\documentclass[a4paper,11pt,article]{article}
\usepackage{textcomp}
\usepackage[english]{babel}
\usepackage{adjustbox} 

\usepackage{graphicx}
\usepackage{booktabs}
\usepackage{graphics}
\usepackage{subcaption}
\usepackage{mwe}
\usepackage[utopia]{mathdesign}
\usepackage{bbm}
\usepackage{booktabs}
\usepackage{dcolumn}
\usepackage{dsfont}
\usepackage{setspace}
\usepackage{comment}
\usepackage{tikz}
\usepackage{pgfplots}
\pgfplotsset{width=10cm,compat=1.9,title style={at={(.35,-.28)}}}
\usetikzlibrary{calc}
\usepackage{float} 
\usepackage{mathpazo} 
\usepackage[T1]{fontenc}
\normalfont

\usepackage{amsthm}
\usepackage{amsmath}
\usepackage{gensymb}
\usepackage{wasysym}
\usepackage{marvosym}
\usepackage[margin=1 in]{geometry}
\usepackage{tikz}
\usetikzlibrary{positioning}
\usepackage[titletoc,title]{appendix}
\usepackage{longtable}
\usepackage{multirow}
\usepackage{lscape}
\usepackage{bigstrut}
\usepackage[sort&compress]{natbib}
\usepackage{longtable}

\theoremstyle{plain}

\theoremstyle{definition}
\newtheorem{fnd}{Finding}
\theoremstyle{definition}
\newtheorem{hyp}{Hypothesis}

\theoremstyle{definition}

\usepackage{hyperref}
\hypersetup{
    colorlinks=true,
    linkcolor=blue,
    citecolor=blue,
    urlcolor=cyan,
    pdftitle={Sharelatex Example},
    bookmarks=true,
    breaklinks=true
}

\usepackage{enumitem}

\begin{document}
\newgeometry{top=0.1cm} 

\title{Position Uncertainty in a Sequential Public Goods Game: An Experiment}
\author{Chowdhury Mohammad Sakib Anwar\footnote{Business School, Faculty of Business \& Digital Technology,
University of Winchester, SO22 4NR, Winchester, U.K., {\Letter}: sakib.anwar@winchester.ac.uk, ORCID: 0000-0002-0223-7963}
\and
Konstantinos Georgalos\footnote{Department of Economics, Lancaster University Management School, LA1 4YX, Lancaster, U.K.,{\Letter}: k.georgalos@lancaster.ac.uk, ORCID: 0000-0002-2655-4226\newline
We are grateful to the Department of Economics from the Lancaster University Management School for providing the funds for this study. We are thankful to Renaud Foucart, Alexander Matros, Sonali Sen Gupta, David Rietzke and participants at the 2023 Liverpool Economic Theory and  Experimental Workshop, the ESA European Meeting at Exeter and the 2023 Newcastle Experimental Economics Workshop. The study has been pre-registered  and details, as well as instructions, data and replication files can be found at \url{https://osf.io/br5uy} \newline\textbf{Conflict of Interest}: None }.
}

\maketitle

\begin{abstract}
\setstretch{1.5}
\citet{gallice2019co} present a natural environment that sustains full co-operation in one-shot social dilemmas among a finite number of self-interested agents. They demonstrate that in a sequential public goods game, where agents lack knowledge of their position in the sequence but can observe some predecessors' actions, full contribution emerges in equilibrium due to agents' incentive to induce potential successors to follow suit.
In this study, we aim to test the theoretical predictions of this model through an economic experiment. We conducted three treatments, varying the amount of information about past actions that a subject can observe, as well as their positional awareness.
Through rigorous structural econometric analysis, we found that approximately 25\% of the subjects behaved in line with the theoretical predictions. However, we also observed the presence of alternative behavioural types among the remaining subjects. The majority were classified as conditional co-operators, showing a willingness to cooperate based on others' actions. Some subjects exhibited altruistic tendencies, while only a small minority engaged in free-riding behaviour.


\end{abstract}
\pagebreak

\restoregeometry 

\section{Introduction}

In a wide range of economic settings such as crowdfunding projects \footnote{\label{note1}The Ocean Cleanup's record-breaking crowdfunding campaign raised over 2 million USD, enabling them to advance in their mission to rid the oceans of plastics (See https://theoceancleanup.com/milestones/crowd-funding-campaign/). Governments are increasingly using crowdfunding for conservation funding. Examples include the US National Recreation and Park Association's "Fund Your Park" and Australia's state of Victoria's "Threatened Species Collection" initiative.}, open-source software development, and environmental conservation initiatives, the power of sequential contributions has proven pivotal in fostering collective action and yielding positive outcomes for the common resource. In crowdfunding projects \citep{stiver2015civic,hudik2018private,ansink2022crowdfunding}, the act of individuals contributing money to support a cause sets off a chain reaction where others are motivated to join in, progressively building up the project's funding and increasing its chances of success. Similarly, in open-source software development \citep{athey2014dynamics,xu2020makes,von2003community,lerner2002some}, programmers' contributions, whether in the form of code, bug fixes, or improved documentation, lead to an iterative improvement in the software's quality and functionality, enticing more developers to get involved in the project's advancement. Environmental conservation initiatives (see Footnote \ref{note1} for examples of environmental conservation) also benefit from sequential actions, as initial efforts by one group to preserve natural resources inspire others to participate, resulting in a cumulative and lasting improvement of the environment. In all these examples, the dynamic of sequential contributions fosters cooperation and participation, driving the collaborative spirit towards benefiting  the overall well-being of the community.

In standard social dilemmas with self-interested  players, co-operation is usually hindered by the players' free-riding motives , leading often to socially sub-optimal outcomes. On the other hand, co-operation is still feasible, in the case of a repeated strategic interaction over an infinite horizon, or whenever the players' preferences are non-standard (e.g. altruistic preferences). In a recent study, \citet{gallice2019co} (G\&M henceforth), in the context of a discrete public goods game, show that it is possible  to achieve full co-operation, even in the case where the interaction is a one-shot game, the number of players is finite, and the agents are self-interested.

In this model, individuals have to make decisions sequentially, without knowing their position in the sequence (position uncertainty), but are aware of the decisions of some of their predecessors by observing a sample of past play. In the presence of position certainty, those placed in the early positions of the sequence would want to contribute, in an effort to induce some of the other group members to co-operate \citep{rapoport1994provision}, while late players, would want to free-ride on the contributions of the early players. Nevertheless, if the agents are unaware of their position in the sequence, they would condition their choice on the average payoff, from all potential positions, and they would be inclined to contribute  so as to induce the potential successor to do so as well. G\&M show that full contribution can occur in equilibrium, where given appropriate values of the parameters of the game (i.e. return from contributions), it is predicted that there exists an equilibrium where all agents contribute.

The main intuition behind the model is that an agent who observes a sample of past decisions of immediate predecessors, without defection, would decide to contribute, hoping to influence all her successors to do so. If instead, she decides to defect, then all the successors are expected to defect as well. As there is position uncertainty, the agent deals with a trade-off between inducing contributions from the remaining players in the sequence and the cost of contributing. In their main theoretical result, G\&M  show that incentives off the equilibrium path, largely depend on the sample size that the agents observe. When an agent observes a sample of more than one previous actions that contains defection, there is no way this agent can prevent further defection by choosing to contribute. On the other hand, when the sample size is equal to one (only the decision of the previous player is observed), the agent can induce further contributions from the remaining agents in the sequence, by deciding to contribute. The model predicts that when the sample size is equal to  one there is a mixed  strategy equilibrium  that can lead to full contribution. Finally, when the agents are aware of their position in the sequence, the model predicts that full contribution will unravel, as late agents in the sequence will have an incentive to free-ride.

In this paper, our objective is to contribute to the understanding of agents' decision-making process when they are presented with partial information about past contributions and face uncertainty about their position in the sequence. Motivated by the G\&M theoretical contribution, the objective of our study is threefold, and in particular, we aim to identify: (1) How does the content and the size of past history, regarding contribution actions, affects contributions in the sequential public goods game when there is position uncertainty, (2) What is the effect of positional awareness 
on contribution decisions, when the agents receive partial information on past actions, and (3) What reasons lead to larger group contributions, when the theory predicts defection, or in other words, what is the role of agents with altruistic motives in explaining our data. To address these questions, in an incentivised economic experiment, with three different treatments,
we vary the amount of information (size of sample of past actions that a subject can observe)
as well as the positional awareness that subjects have.

We design a 2$\times$2 fractional factorial experiment to test decision-making in sequential public goods games under varying levels of information and sequence awareness. Each session involved ten rounds of a sequential public goods game. Participants were randomly re-matched in each round into new groups of four and tasked with allocating tokens in a common project or private account. The treatments tested the impact of observing a sample of past decisions—ranging from one to two immediate predecessors—on subsequent contribution choices, without complete knowledge of one's position in the sequence, except for one treatment where participants' sequence positions were disclosed.  Using the \citet{Selten1967} \textit{strategy response} method, we collected comprehensive data on participant decisions across all possible information scenarios they might encounter.

 We find that while participants' behaviour is in line with the theoretical predictions, there is still a large part of behaviour that the model cannot account for. Using the Strategy Frequency Estimation Method \citep{dalbo11,fudenberg12}, we allow for the presence of  various behavioural types in our subject population,  and we estimate the proportion of each type in our data (see \citealt{fischbacher2001people}; \citealt{bardsley2007experimetrics}; \citealt{thoni2018conditional}; \citealt{katuscak2023drives}; \citealt{preget2016}). On top of G\&M agents, we classify subjects to  \textit{free-riders} (who never contribute), \textit{altruists} (who always contribute no matter their position), and \textit{conditional co-operators} (who always contribute if they are in position 1 and contribute if at least one other person in the sample has contributed when they are in positions
2-4). Additionally, we investigate whether subjects align with the predictions of the G\&M model (\textit{G\&M type}). We find that around 25\% of the subjects behave according to the G\&M model, the vast majority behaves in a conditional co-operating or altruistic way, and a non-significant proportion free rides. From a mechanism design point of view, we find that introducing uncertainty regarding the position, along with a constrained sample of previous actions (i.e. only what the immediate precedent player), maximises the public good provision.

Our work is related to and extends various strands of the literature, which we  briefly summarise below. Prior to G\&M's research,  the timing of contributions and the level of funds raised had received considerable attention in the theoretical literature. \citet{Varian1994} shows that, under appropriate assumptions, a sequential contribution mechanism elicits lower contributions than a simultaneous contribution mechanism.  The crux of this result lies in the set-up of the model, where a first mover may enjoy a first-mover advantage and free-ride. On the other hand, \citet{cartwright2010} using a sequential public goods game with exogenous ordering, show that agents early enough in the sequence would want to contribute, if they believe that imitation from others is quite likely. In the context of fundraising, \citet{romano2001charities} examine the conditions under which a charity prefers to announce contributions in the form of a sequential-move game, while \citet{vesterlund2003informational} shows that an announcement strategy of past contributions, not only helps worthwhile organisations to reveal their type, but it also helps the fundraiser reduce the free-rider problem, a result that  \citet{potters2005after} confirm experimentally.

The early and recent experimental literature has provided substantial evidence on the superiority of a sequential contribution mechanism compared to a simultaneous one (see \citealt{erev1990provision}; \citealt{rapoport1994provision}; \citealt{rapoport1997}, in step-level public goods games; and
\citealt{Andreoni2002}; 	\citealt{coats2009simultaneous}; \citealt{gachter2010sequential}; \citealt{figuieres2012vanishing}; \citealt{teyssier2012inequity} in public goods games without a threshold). The vast majority of the aforementioned studies conclude that the sequential protocol is significantly more effective in solving the public goods problem, compared to the simultaneous protocol, and that the effect of information on contribution is dramatic \citep{erev1990provision}. \citet{suleiman1994position} highlight two properties that define the sequential protocol: (1) information regarding the position in the sequence, and (2) information about the actions of preceding players. Therefore, there may be various information structures that underlie social dilemmas or the provision of a public good in real life, and relaxing or modifying these structures can lead to more realistic protocols of play. Most of the previous literature has focused on the comparison between the simultaneous and the sequential mechanism, while in terms of available information, the most commonly employed information structure is either full information on past decisions or no information at all. Both features appear to be closer to the reality that characterises the provision of public goods. In our experiment, rather than comparing the effectiveness of different contribution mechanisms,  our focus is on the effect of the  available information provided to the subjects, as well as the effect of positional awareness on the contribution choices. Table \ref{tab:literature} summarises the experimental design features of the studies closest to ours.

A similar result regarding the superiority of the sequential mechanism has also been established in the literature on general public goods, particularly in the context of common pool resource games \footnote{Sequential mechanisms have also been analysed in give-some and take-some social dilemma games, for example, see \citet{tung2013sequential}.}. This literature has identified significant ordering effects, even in the case where later subjects in a sequence could not observe past decisions (see \citealt{rapoport1993sequential}; \citealt{budescu1995positional}; \citealt{suleiman1996fixed};  \citealt{rapoport1997}). The model we test, makes sharp predictions regarding the behaviour of the agents for  each of the potential positions in the sequence, subject to the available size and content of the sample of past decisions. Our experimental design allows us to manipulate the content of this sample, in terms of the number of past contributions, and observe the role of this information in this kind of ordering effect. While the main theoretical prediction in step-level public goods games is that the players at the early positions of the sequence will free-ride, the framework we are exploring predicts that players at the beginning of the sequence will contribute instead, to incentivise their successors.  This is closely related to the leading-by-example literature, based on the linear public goods game (see \citealt{gachter2012}; \citealt{levati2007leading}; \citealt{potters2007leading}; \citealt{guth2007leading}; \citealt{figuieres2012vanishing}; \citealt{sutter2014leadership}; \citealt{preget2016}, in the context of public goods games, or \citealt{moxnes2003effect}, in a public bad experiment). They all find robust evidence of first movers contributing more than later movers, and later movers' contributions to be positively correlated to the first movers' contributions, indicating reciprocal motives.

Directly linked to the last observation, \citet{figuieres2012vanishing} highlight that many individuals condition their decisions on the observation of others' decisions in finitely repeated simultaneous public goods games (see \citealt{fischbacher2010}; \citealt{keser2000conditional}). Extensive experimental research in public goods games has shown that a large proportion of  subjects' behaviour deviates from the predictions of Nash equilibrium and they mostly behave in a \textit{conditionally} co-operating manner. The latter implies that their contribution is positively correlated to their beliefs about the contributions of the other members of the group. \citet{fischbacher2001people} was among the first studies to classify subjects to different \textit{types} of decision makers, including the \textit{selfish} type (free-rider), the \textit{reciprocator} type, and the type applying a \textit{hump-shaped} strategy (contributions that are first increased then decreased in others' contributions). This methodology was later extended in \citet{bardsley2007experimetrics}, refined in  \citet{thoni2018conditional}, and recently used in \citet{katuscak2023drives}, while \citet{preget2016} use this in a sequential leader-followers public goods game. In addition to testing the predictions of G\&M, we use structural econometric modelling to test the presence of alternative behavioural types of agents in our data. In particular, using the Strategy Frequency Estimation Method (SFEM), an estimation procedure introduced in \citet{dalbo11} and \cite{fudenberg12}, we explore the presence of free-riders, altruists and conditional co-operators in our data, on top of those behaving as the G\&M model prescribes.

Contributing to the literature on uncertainty in social dilemmas \citep{suleiman2001provision,van2004we,budescu2002model,au2003effects}, this paper examines how different types of uncertainty impact cooperative behaviour. Studies like \citet{suleiman2001provision} and \citet{au2003effects} show varying effects: threshold uncertainty can increase contributions under certain conditions, while group size uncertainty paradoxically enhances cooperation. These findings, contrasting with the non-cooperative behaviour often induced by resource size uncertainty, align with our observations of increased cooperation under positional uncertainty. The nature of positional uncertainty, unlike the direct impact of resource size uncertainty on resource availability, may not readily justify non-cooperative behaviour. Our research, corroborating the significance of position in sequential decisions highlighted by \citet{budescu1995positional,budescu2002model}, adds to the understanding of how context influences cooperative dynamics in social dilemmas.

In our study, the role of positional uncertainty parallels the findings of \citet{au2003effects} on group size uncertainty, where certain uncertainties promote cooperative behaviour. This contrasts with \citet{rapoport1993sequential}'s observations of non-cooperative behaviour under resource size uncertainty. Such distinctions underscore the complex impact of uncertainty types: positional and group size uncertainties might foster cooperation through a collective sense of ambiguity, while resource size uncertainty, directly affecting perceived resource availability, can lead to competitive behaviour.

The rest of the paper is organised as follows. In section \ref{sec:theory} we present the main theoretical predictions of \citet{gallice2019co}. Section \ref{sec:design} presents the experimental design and the procedures, while section \ref{sec:results} presents the descriptive statistics and results of the structural econometric analysis. We then conclude in section \ref{conclusion}.
\begin{landscape}

\begin{table}[]
\begin{adjustbox}{scale=0.8}
\begin{tabular}{@{}lcccccccccc@{}}

\toprule
Study                           & Group size & Rounds & Contribution & Threshold & Repeated & Position & \begin{tabular}[c]{@{}c@{}}Information on\\ past contribution\end{tabular} & Different types  &  \begin{tabular}[c]{@{}c@{}} No. of \\ Subjects \end{tabular}  &\begin{tabular}[c]{@{}c@{}} Payment \\ Method \end{tabular}\\ \midrule
\citet{erev1990provision}      & 5          & 1      & Binary       & Y       & N       & known    & full                                                                         & N               & 75 &Total\\
\citet{erev1990provision}      & 5          & 1      & Binary       & Y       & N       & unknown  & \#past contributors                                                                           & N               & 60&Total\\
\citet{erev1990provision}      & 5          & 1      & Binary       & Y       & N       & unknown  & \#past defectors                                                                      & N               & 60&Total\\
\citet{rapoport1994provision}  & 5 or 7          & 1      & Binary       & Y       & N       & known    & full                                                                         & N               & 70&Total\\
\citet{rapoport1994provision}  & 6 or 7          & 1      & Binary       & Y       & N       & unknown  & \#past contributors                                                                           & N               & 70&Total\\
\citet{rapoport1994provision}  & 6 or 7          & 1      & Binary       & Y       & N       & unknown  & \#past defectors                                                                      & N               & 70&Total\\
\citet{Andreoni2002}           & 2          & 14     & Continuous   & N        & N       & known    & full                                                                         & N               & 42&Total\\
\citet{coats2009simultaneous}  & 4          & 20     & Continuous   & Y       & N       & known    & previous                                                                      & N               & 108&Total\\
\citet{gachter2010sequential}  & 2          & 15     & Continuous   & N        & N       & known    & full                                                                         & N               & 128&Random\\
\citet{teyssier2012inequity}   & 2          & 1      & Continuous   & N        & N       & known    & full                                                                         & Y              & 118&Random\\
\citet{figuieres2012vanishing} & 4 or 8     & 15     & Continuous   & N        & Y      & known    & full                                                                         & Y              & 72&Total\\
\citet{figuieres2012vanishing} & 4 or 8     & 15     & Continuous   & N        & Y      & known    & unknown                                                                       & Y              & 80&Total\\
this study                      & 4          & 10     & Binary       & N        & N       & unknown  & 1 previous                                                                    & Y              & 32&Random\\
this study                      & 4          & 10     & Binary       & N        & N       & unknown  & 2 previous                                                                    & Y              & 32&Random\\
this study                      & 4          & 10     & Binary       & N        & N       & known    & 2 previous                                                                    & Y              & 32&Random\\ \bottomrule
\end{tabular}
\end{adjustbox}
\caption{{\small Summary of the experimental design of previous studies on sequential public goods games. Contribution: Continuous or 'all-or-nothing'. Threshold: Threshold or standard public goods game. Repeated: One-shot or repeated game (with strangers' matching protocol). Different types: Exploration of alternative behavioural models. Payment Method: Total rounds or a randomly chosen subset.
All experiments involved financially motivated participants from student subject pools.}}

\label{tab:literature}
\end{table}

\end{landscape}

\section{Theoretical Framework}\label{sec:theory}
\citet{gallice2019co} propose a natural environment that sustains full-co-operation in one-shot social dilemmas among a finite number of self-interested agents, in the context of a public goods game. Here we present the main features and predictions of the model and we refer the interested reader to the original study for further details. In an effort to be as consistent as possible with the original study, we adopt the same notation as the authors. In summary, the main prediction of the model is that contributions from all individuals in a group will emerge when agents make decisions sequentially, do not know their position in the sequence, and observe the decisions of some of their predecessors.

Let $I=\{1, \ldots, n\}$ be a set of risk-neutral agents making choices in a sequence.  They know the length $n$ of the sequence but are uncertain about their position. Players are exogenously placed in the sequence that determines the order of play and they can be at any position with equal chances. Therefore, they have symmetric position beliefs regarding their position in the sequence. When the turn of play for agent $i \in I$ arrives, she can observe a sample of her predecessors' actions. She then chooses one of two actions  $a_i \in\{C, D\}$, with $a_i=C$ being contribution of a fixed amount 1 to a common pool, while $a_i=D$ denotes defection, and therefore, investing $0$. After all players choose an action, the total amount invested gets multiplied by the return from contributions parameter $r$, and is equally shared among all agents.
Let $G_{-i}=\sum_{j \neq i} \mathbb{1}\left\{a_j=C\right\}$ denote the number of agents in the group who contribute, so $G_{-i}$ can take any value in the set $\{0, \ldots, n-1\}$. The payoffs $u_i\left(a_i, G_{-i}\right)$ from contributing or defecting can thus be expressed as:
\[
u_i\left(C, G_{-i}\right)=\frac{r}{n}\left(G_{-i}+1\right)-1 \quad \text { and } \quad u_i\left(D, G_{-i}\right)=\frac{r}{n} G_{-i} .
\]
A value of $r$ that satisfies $1<r<n$, guarantees that although contribution by all agents is socially optimal, each agent strictly prefers to defect for any given $G_{-i}$, as is the case in the standard public goods game.

Now we introduce the idea of \textit{sampling}. Before choosing an action, each player observes how many of her $m\geq 1$ predecessors contributed. Agents in positions 1 to $m$ observe less than $m$ actions, given that there are less than $m$ players before them. For instance, if $m=2$, the player at position 1 observes a null sample, the player at position 2 observes a sample of only one previous action, while the player at position 3 observes a sample of two past actions. Each sample is denoted as a pair:
\[\xi=(\xi', \xi'')\]
where the first component indicates the number of agents sampled, and the second component is the number of agents who contributed, in that sample.\footnote{For instance, $\xi=(2,2)$ indicates a sample of full contribution by the previous 2 players,   $\xi=(2,0)$ indicates a sample of the two previous players where none contributed, and  $\xi=(2,1) $ indicates a sample of the two previous players, where only one of the two contributed. The first player in the sequence always receives the sample $\xi=(0,0)$.}.The agent has no way to identify the action of each player, but only the size of the sample and the total contribution. As a result, the first $m$ agents can infer their position given the sample they observe. According to this set-up, players face an extensive-form game with incomplete information, and the solution concept is based on \citet{kreps1982} notion of \textit{sequential equilibrium}. Let  $\Xi$ the set of all samples that an agent can observe. The  strategy of player $i$ is a function:
\[\sigma_i(C \mid \xi): \Xi \rightarrow[0,1]\]
that specifies the probability of contributing given the sample that the agent observed.\footnote{Let $\sigma=\left\{\sigma_i\right\}_{i \in I}$ denote a strategy profile and $\mu=\left\{\mu_i\right\}_{i \in I}$ a system of beliefs. A pair $(\sigma, \mu)$ represents an \textit{assessment}. Assessment $\left(\sigma^*, \mu^*\right)$ is a sequential equilibrium if $\sigma^*$ is sequentially rational given $\mu^*$, and $\mu^*$ is consistent given $\sigma^*$. Belief formation is needed in order for the agent to find out her position in the game and also about the history of actions that led to this position.}

G\&M characterise an equilibrium where all the agents contribute adopting a simple profile of play, namely, agents contribute unless they observe a defection.  In particular,  their Proposition 1 \citep[p. 2142]{gallice2019co} states that if:
\begin{equation}\label{eq:prop1}
r\geq 2\left(1+\frac{m-1}{n-m+1}\right)
\end{equation}
then, there exists an equilibrium in which all agents in the sequence contribute. To prove this proposition, G\&M derive a series of lemmas, which our experiment aims to test. We focus on three particular cases: (1) when the sample size is greater or equal to 2, (2) when the sample size is equal to 1, and (3) when there is position certainty.
\subsection{Sample $m \geq 2$}
Let $\Xi^C$ be the set of all samples without defection. Then, Lemma 1 states that, assuming the following profile of play for all \(i \in I \):
\begin{align*}
    \sigma_i^*(C\mid \xi) =
\begin{cases}
1, & \text{if $\xi \in \Xi^C$ } \\
0, & \text{otherwise.}
\end{cases}
\end{align*}
it follows that $(\sigma^*,\mu^*)$ is a sequential equilibrium provided that the inequality in expression \ref{eq:prop1} is satisfied. In other words, what the above expression states, is that a player will always contribute, unless she observes at least one defection.
A sketch of the intuition behind this follows. Consider an agent who observes a sample of full co-operation. According to the equilibrium definition, this occurs at the equilibrium path, and the agent can therefore infer that all the previous agents in the sequence contributed. By contributing herself, she knows that all subsequent players will contribute as well, and therefore, her expected payoff from contributing will be equal to $r-1$, and this payoff is independent of the agent's beliefs regarding her position in the sequence. On the other hand, the payoff from defecting depends on these beliefs. Consider an agent who observes a sample of $m$. This agent can therefore infer that she is not placed in the first $m$ positions, and there are equal chances of her being in any of the positions in $\{m+1,\cdots, n\}$. The expected position is $(n+m+1)/2$, which means that she can expect that $(n+m-1)/2$ agents have already contributed. The payoff from defecting is then equal to $(r/n)(n+m-1)/2$. Combining the two, gives the condition that $r$ needs to satisfy in order for the agents to be willing to contribute, whenever they observe full contribution samples (condition \ref{eq:prop1}).

On the other hand, the equilibrium profile requires that an agent should defect, upon observing a sample with at least one defection. G\&M provide arguments why an agent who observes defection cannot affect her successors' actions, regardless the value of $r$, and she is better off defecting herself as well.\footnote{The main argument is that an agent should defect for three reasons. Consider the case of $m=2$. First, observing the sample (2,0), means that no matter her choice, the subsequent agent will observe a sample with at least one defection. Second, observing a sample (2,1), the agent has no way to identify whether the defection took place at position $t-1$ (her immediate predecessor), or $t-2$ (the one before). In the former case, a same argument as before applies. In the latter one, it appears that two mistakes happened, that is the agent at position $t-2$ defected, which should not happen at equilibrium, and, the agent at position $t-1$ contributed, after observing a defection.} This leads to our first hypothesis\footnote{For exposition reasons, the numbering of the hypotheses in the paper slightly differs from the one in the preregistration. Hypothesis 1 in the paper corresponds to hypotheses 1 and 2 in the preregistration, Hypothesis 2 in the paper corresponds  to hypothesis 4 in the preregistration, while hypothesis 3 is the same in both the paper and the preregistration.}:
\begin{hyp}\label{hyp:hyp1}
There will be full contribution when there is position uncertainty, and  the agents observe samples $m\geq 2$ of full contribution, compared to the case where agents observe at least one defection.
\end{hyp}
\subsection{Sample $m=1$}
A separate analysis is required for the situation in which only one previous player is sampled. In the scenario where $m=1$, a player who spots defection has the ability to prevent additional defections by deciding to contribute. If the return from contributions is excessively high, the simple strategy of contributing, unless a defection is observed, cannot establish equilibrium. Players would deem it beneficial to contribute after witnessing a contribution but would not consider it advantageous to defect upon observing a defection. When $r$ is  high, a player who notices a defection finds the cost of contribution worthwhile to motivate all her successors to contribute. Consequently, she would choose to contribute rather than defect. However, as $r<n$, the strategy profile in which all players contribute cannot be an equilibrium, because players who witness contribution would deviate to defecting. This means a \textit{pure strategy} equilibrium with total contribution is impossible.

When the multiplication factor $r$ is high, full contribution can emerge in a \textit{mixed strategy} equilibrium. G\&M construct a profile of play where players respond to contribution with contribution and permit a defection to be 'forgiven' with a probability $\gamma \in[0,1)$. The potential for future forgiveness makes defecting more appealing: successors could revert to contribution. This leads to  Lemma 2 which states:
\begin{align*}
    \sigma_i^*(C\mid \xi) =
\begin{cases}
1, & \text{if $\xi \in \{(0,0),(1,1) \}$ } \\
\gamma, & \text{if $\xi=(1,0)$.}
\end{cases}
\end{align*}
for  $r \in \left(3-3/(n+1), n \right) $, that is agents always contribute when they observe  contribution from their immediate predecessor and contribute with probability $\gamma$ if they observe defection. Our second hypothesis is summarised as follows:
\begin{hyp}\label{hyp:hyp2}
There will be full contribution when there is position uncertainty, and  the agents observe sample $m=1$ of full contribution. In addition, there will be a mixed strategy equilibrium, with mixing probability $\gamma$, when the agents observe sample $m=1$ of defection.
\end{hyp}
\subsection{Position certainty}
\begin{figure}[h!]
    \centering
\begin{tikzpicture}
\begin{axis}[
    title= {},
    axis lines = left,
    xlabel style={at={(axis description cs:1.05,0.04)}},
    xlabel = $t$,
    ylabel = {Payoff},
    ymin=0, ymax=2.5,
    xmin=0, xmax=5,
    legend pos=south east,
    ytick=2, ymajorgrids, xmajorgrids
]
\addplot[domain=0:5, samples=100, color=red, solid]{(3)/4*(x-1)};\addlegendentry{Payoffs from defecting}
\addplot[domain=1:6, samples=100, color=blue, densely dotted, dash pattern=on 3pt off 3pt, very thick]{2}; \addlegendentry{Payoffs from contributing}
\end{axis}
\end{tikzpicture}
    \caption{Payoffs conditional on position and on samples without defection ($r=3, n=4$).}
    \label{fig:unravel}
\end{figure}
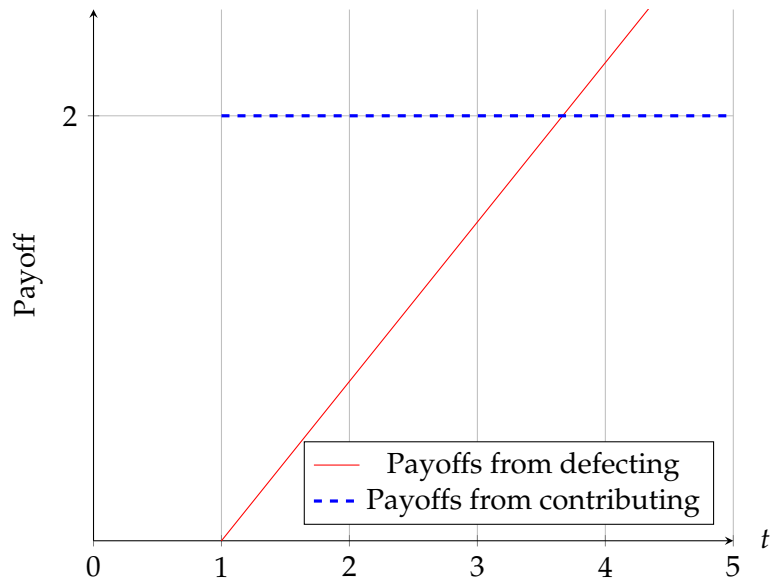

The last theoretical prediction of the model that we test, prescribes that when agents are aware of their position in the sequence, contribution unravels. Assume that player $i$ knows her position, i.e., she knows that she is in position $t$. If all of her predecessors contributed, and she does not contribute, then none of her successors will contribute. This means that exactly $t-1$ players contributed. The payoff from defecting is $(r / n)(t-1)$, while the payoff from contributing is equal to $n-1$, as before. Figure \ref{fig:unravel} illustrates agent \(i\)'s payoffs as a function of her position. The payoff from contribution, for agents early in the sequence, is larger than that of defecting. But agents placed late in the sequence would prefer defection.  Consequently, if agents knew their position,  contribution would unravel.
Therefore, our last hypothesis states:
\begin{hyp}\label{hyp:hyp3}
Full contribution will unravel when the agents observe samples $m=2$ of full contribution and they are aware of  their position in the sequence.
\end{hyp}
\section{Experimental Design and Procedures}\label{sec:design}
To test the predictions of the model presented in the previous section, we designed and conducted an incentivised economic experiment.  The experiment took place at the Lancaster Experimental Economics Lab (LExEL) in February 2023, involving 96 subjects across three treatments\footnote{The power analysis, conducted with a significance level of 0.05 and aiming for 80\% power, to detect a moderate effect size (Cohen's h = 0.65) in the specified one-sided test, indicated a minimum sample of 29.3 subjects per treatment. The power analysis was conducted using the \textit{pwr.2p.test} function from the library \textit{pwr} in R. }. The subjects were primarily Lancaster University undergraduates (82.2\%) from various disciplines, mainly Business and Economics (60\%), Social Sciences (23\%) and Science and Medicine (17\%)\footnote{A potential criticism could be that using a student subject pool could corroborate the results since students tend to be more socially connected compared to a more representative population. Nevertheless, a common finding in the literature is that students tend to behave in a less prosocial way compared to representative populations or professionals (see for example \citealt{anderson2013self}; \citealt{bellemare2007representative}; \citealt{belot2015comprehensive}), giving less in public goods experiments (\citealt{gaechter2004trust}; \citealt{carpenter2011social}) or behaving in a similar manner to non-student populations (\citealt{exadaktylos2013experimental}). 
}. In terms of gender, the sample comprised 48\% female and 52\% male participants. The average age was 20.7 for those who provided details (4 subjects did not reply this question). Recruitment took place  via the ORSEE system \citep{greiner2015subject}, and the  experiment was computerised using the oTree platform \citep{chen2016otree}. None of the subjects had previously participated in a public goods experiment and every subject participated only once.

Each session consisted of 16 participants, forming groups of 4 and playing for a total of 10 rounds. In an effort to replicate the one-shot play environment, we adopted a random matching protocol where in every round, participants were randomly matched into new groups.  At the beginning of all the sessions, participants were given written instructions followed by a comprehension test. Participants could retake the questionnaire until they passed. The interface would provide immediate feedback and explanation on the questions that the subjects failed to respond correctly. An experimenter would offer additional clarifications if there were further questions. Therefore, no exclusion criteria were applied. In every session, participants played a variation of a sequential, binary public goods game, as described in the previous section. At the beginning of every round, participants were endowed with 10 tokens each, with an exchange rate of \pounds 0.50 per token. In every round, subjects were randomly positioned in the sequence (with equal chances of being allocated at each position) and were sequentially asked to make a binary decision between investing all of these tokens in either a common project account or a private account. At the end of every round, feedback was provided regarding the participant's decision, the total group contribution, the individual share, and the payoff from the round. We employed the random payment mechanism, with one of the 10 rounds being randomly chosen to be played for real (different for each group)\footnote{The utilisation of both Experimental Currency Units (ECU) and the Random Payment Mechanism (RPM) has been a subject of ongoing debate in the literature. \citet{drichoutis2015veil} show that there is no difference between using ECUs or cash in the lab; while at the same time, the use of ECUs leads to decisions closer to theoretical predictions. Conversely, empirical research comparing the different incentive mechanisms is inconclusive \citep{azrieli2018incentives}, while \citet{azrieli2020incentives} provide a theoretical justification that the RPM is incentive-compatible in almost all experiments. We follow the common practice that the literature in public goods adopts and we use both.
}. The average payment was \pounds 17.8, including a show-up fee of \pounds 5. The experimental sessions lasted less than 45 minutes and payments were made via bank transfer.

The main objective of our experiment was to test the predictions of G\&M when two dimensions change, namely, the size of the sample of past decisions observed by a subject, and whether subjects are aware of their position in the sequence. We, therefore, have a 2$\times$2 fractional-factorial design, where the sample size can be either 1 or 2 past actions, and the position can be unknown or known.\footnote{From the factorial design we dropped the treatment with a sample of \(t-1\) and position certainty as it is a subset of the treatment with a sample of 2 and position certainty and does not bring any particular theoretical interest.}

\begin{figure}[h] \centering
\includegraphics[scale=.75]{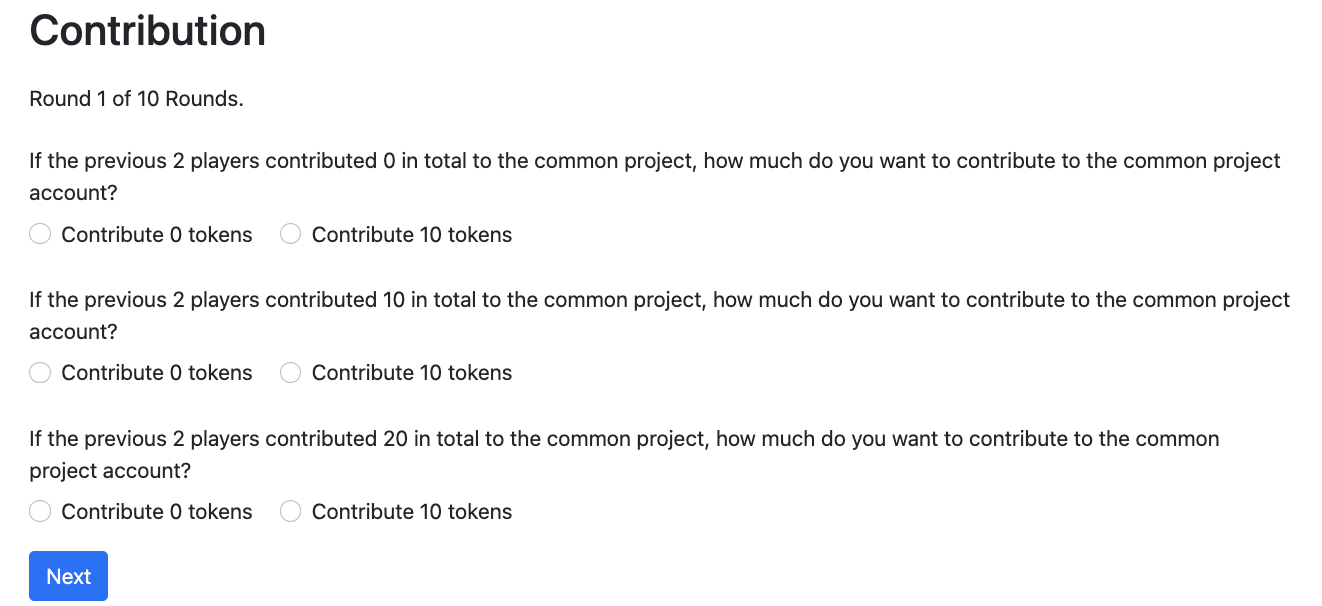}
\caption{Screen shot from Treatment 1 (players in position 3 or 4).}
\label{fig:screenshot}
\end{figure}

There are in total three treatments. Treatment 1 ($T_1$) was designed to test Hypothesis \ref{hyp:hyp1}. In that treatment, participants received a sample  of past decisions of length equal to two (they would receive information on what the combined contribution of the immediate 2 predecessors was\footnote{In line with the theoretical model, the subject could only observe whether the combined contribution of the two previous players was equal to 0, 10 or 20. That is, by observing a sample of length 2 with total contribution 10, the subject could infer that one of the previous two players defected, but had no way to tell which one.}) and  also,  participants were not informed of their position in the sequence unless they were Player 1 or Player 2\footnote{Subjects at positions 1 and 2 could infer their position, given their observed sample was of length 0 and 1 respectively.}. To rigorously test the predictions of the model, we need information on what subjects' choices are, in all potential samples they can observe (i.e. the combined contribution they observe). In particular, the model predicts that if a player receives a full contribution sample (i.e. the combined contribution of the two previous players is equal to 20), she will contribute herself, otherwise, if the contribution is 0 or 10, she will defect. We elicit behaviour using the \citet{Selten1967} \textit{strategy response} method  and collect the participants' choices for each possible scenario. This means participants were asked to make conditional decisions for each possible information set depending on their randomly assigned position in the sequence. While this can be seen as a potential limitation of our design, this elicitation method was necessary to allow us to elicit the full strategy profile of the participants, and therefore, to be able to test the theoretical predictions of the model.\footnote{The strategy method has been extensively used in the framework of public goods games (see \citealt{bardsley2000control}; \citealt{fischbacher2001people}; \citealt{kocher2008conditional}; \citealt{herrmann2009measuring}; \citealt{fischbacher2010}; \citealt{teyssier2012inequity};  \citealt{martinsson2013}; \citealt{katuscak2023drives}), while previous studies found no statistical differences in subjects' responses between the strategy and the direct response method (see \citealt{brandts2000hot};  \citealt{brandts2011strategy}, or \citealt{keser2021strategy} for a discussion).} Figure \ref{fig:screenshot} presents a screenshot from the experimental interface, of what a player in position 3 or 4 could see. In that case, the only information the subject has is that she is positioned in one of the two last positions of the sequence and is asked about her contribution choice, conditional on the potential  total contribution of the two immediate players before her (either of players 1 and 2, if she is player 3, or players 2 and 3, if she is player 4) but she cannot distinguish her position. Therefore, the subject does not know whether there is another player following in the sequence. The available options for players at positions 1 and 2 were adapted accordingly.

Participants were incentivised to reveal their preferences truthfully, as their payment depended on what they and the other members of the group had chosen in that round. In particular, at the end of the experiment (and if this round had been selected for actual play), the computer recalled and matched all participants' decisions in that round and calculated the payoffs based on the scenario that truly transpired. For instance, if the player at position  1 chose not to invest in the common project, the experimental software recalled the decision of the player in position 2, for that specific scenario and so forth, returning the total contribution, the total returns, and the payoffs for each participant in the group.

Treatment 2 ($T_2$), was designed to test Hypothesis \ref{hyp:hyp2}. This treatment mirrored $T_1$, but subjects could observe a  sample of past decisions of size 1 (i.e. the immediate predecessor). Participants were facing position uncertainty, unless they were assigned to be Player 1, for that round. Again, subjects were asked to indicate their contribution decision, conditional on all the potential scenarios, depending on their position in that round. Finally, Treatment 3 ($T_3$), was designed to investigate the impact of position \textit{certainty} on the level of contributions, thereby testing the predictions of Hypothesis \ref{hyp:hyp3}. The features of the treatment are identical to those in $T_1$, with the only difference being that participants were now informed about their role (i.e. position in the sequence). Subjects were asked about their contribution choice, which was conditioned on the potential total contribution of the two immediate predecessors in the sequence. Table \ref{tab:Prediction} summarises the details of our treatments.

\begin{table}[h]
\centering
\resizebox{\linewidth}{!}{%
\begin{tabular}{@{}cccccl@{}}
\toprule
 Treatment  & No. of subjects & Group size& Sample Size & Position & Prediction \\ \midrule
T1 & 32 & 4 & 2              & unknown  & \begin{tabular}[c]{@{}l@{}}Contribute only if $r \geq 2.66$ \\ and observe full contribution.\end{tabular}          \\
T2 & 32 & 4 & 1              & unknown  & \begin{tabular}[c]{@{}l@{}}Pure when $r \in [2,2.4]$. \\ Mixed when $r \in (2.4, 4]$.\end{tabular} \\
T3 & 32 & 4 & 2              & known    & Contribute only if $r \geq 4$.          \\ \bottomrule
\end{tabular}

}
\caption{Summary of all the Treatments}
\label{tab:Prediction}
\end{table}
We conclude this section with a comment on the choice of the value of the parameter  $r$ (the return from contributions parameter). We set the value of  $r$  to be equal to 3, for three reasons: (1) this value of $r$ satisfies the inequality in condition \ref{eq:prop1}, ensuring that subjects prefer to contribute, when they observe full contribution in $T_1$, (2) when $r\in (2.4, 4]$, the model predicts that there exists a mixed strategy equilibrium, with mixing probability $\gamma=0.528$\footnote{ \citet[Appendix p.2150]{gallice2019co} show that $\gamma$ should satisfy the condition:
\[\frac{2}{\gamma}-\frac{(n-1)(1-(1-\gamma)^n)}{\gamma n -1+(1-\gamma)^n}=\frac{n}{r}\]
Solving this numerically, for $n=4$ and $r=3$, we get $\gamma=0.528$, which is the closest we could get to 0.5. $\gamma$ is linearly increasing in the $(2.4,4]$ interval, with $\gamma=0.12$ when $r=2.5$, and $\gamma=1$ when $r=4$.} in $T_2$, and (3) when $r<4$, it is predicted that full contribution will unravel in $T_3$.

\section{Results}\label{sec:results}
First, we present some aggregate results on average distributions along with hypotheses testing. Then, we set up a structural econometric model and we test the presence of  alternative behavioural types in our data, by classifying subjects to various types. We have data from three treatments, where we vary the size of the sample and the position awareness. Given that we are using the strategy method, our data elicit the full strategy profile of each subject, on all potential conditions. That is, depending on the treatment and the position of the subject in the sequence, there are three potential conditions, namely, what subjects would do when they observe a null sample (no one in the sample has contributed), when they observe only one contribution and one defection  ($T_1$ and  $T_3$), and when they observe full contribution. Let us call these conditions $c_0, c_1$ and $c_2$, respectively. Depending on the position of the player in the sequence, subjects with different positions would state different sets of strategies (a subject in position 1 would only state whether she wants to contribute or not, while a subject in the last position would state conditional preferences on all $c_0, c_1$ and $c_2$).

\begin{figure}[htbp]
  \centering
  \begin{minipage}[t]{0.85\textwidth}
    \centering
    \subcaptionbox{Treatment 1}
    {\includegraphics[width=\textwidth,height=0.29\textheight,keepaspectratio]{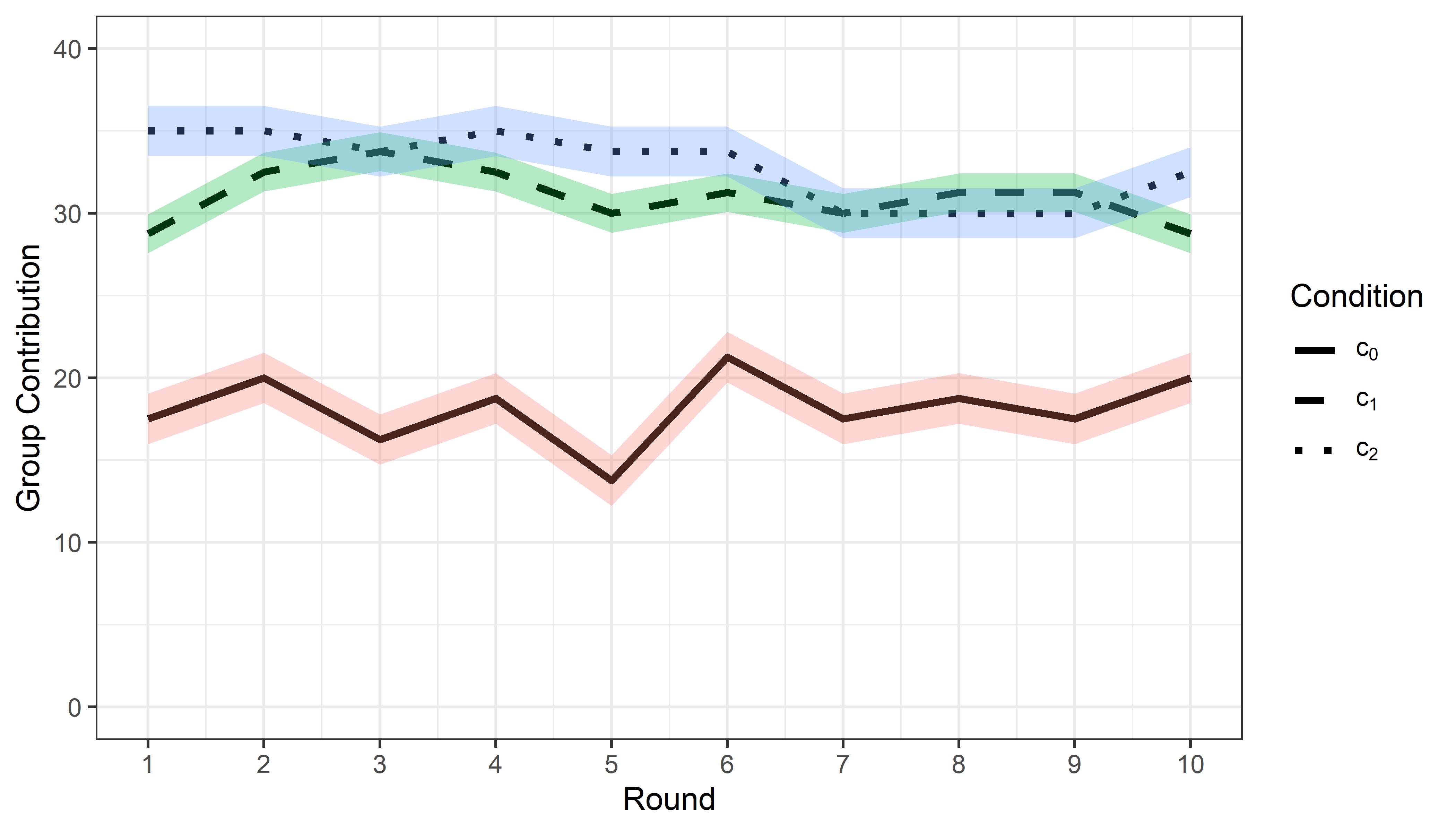}}
  \end{minipage}
  \begin{minipage}[t]{0.85\textwidth}
    \centering
        \subcaptionbox{Treatment 2}
   { \includegraphics[width=\textwidth,height=0.29\textheight,keepaspectratio]{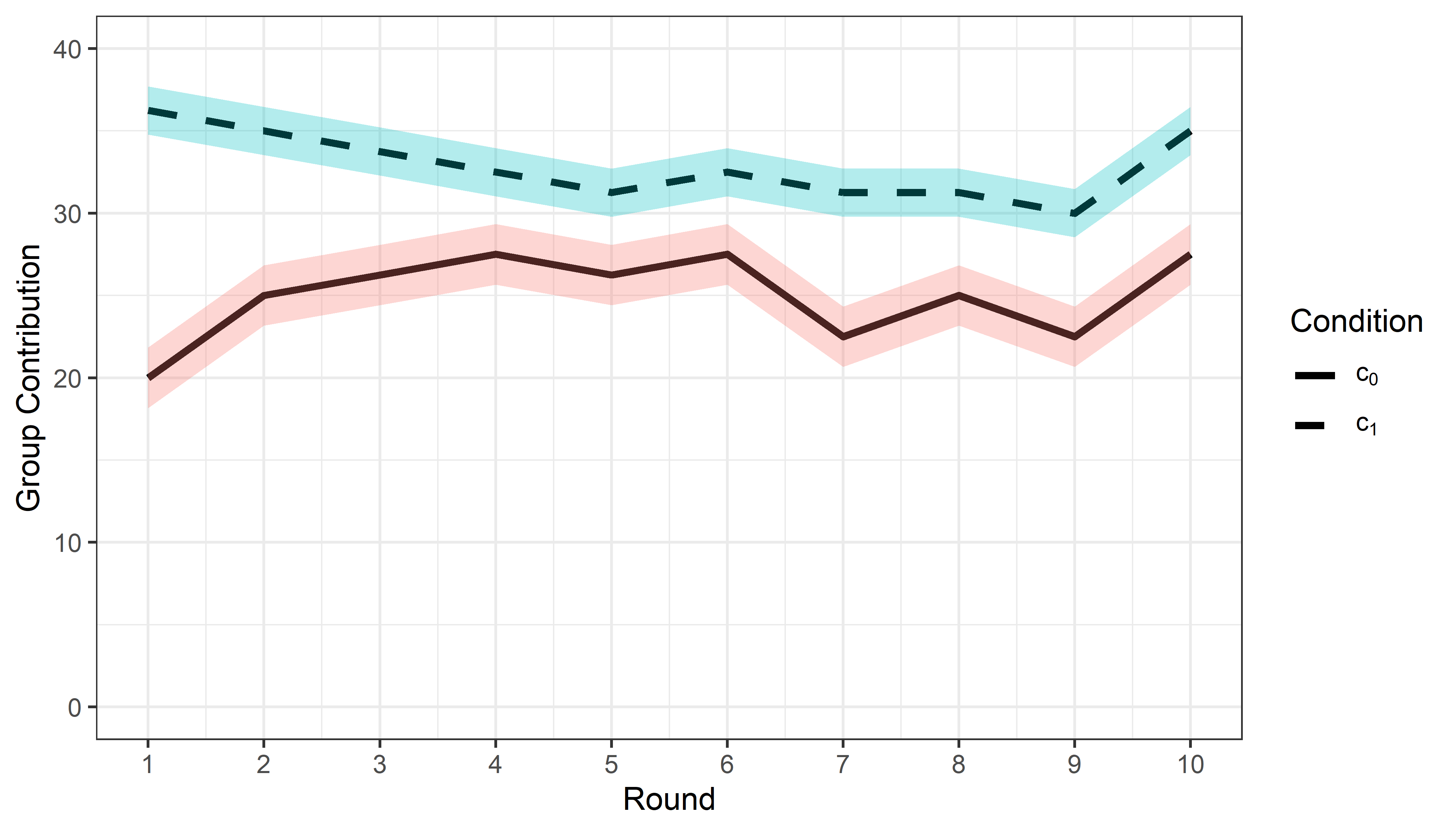}}
  \end{minipage}

  \begin{minipage}[t]{0.85\textwidth}
    \centering
    \subcaptionbox{Treatment 3}
    {\includegraphics[width=\textwidth,height=0.29\textheight,keepaspectratio]{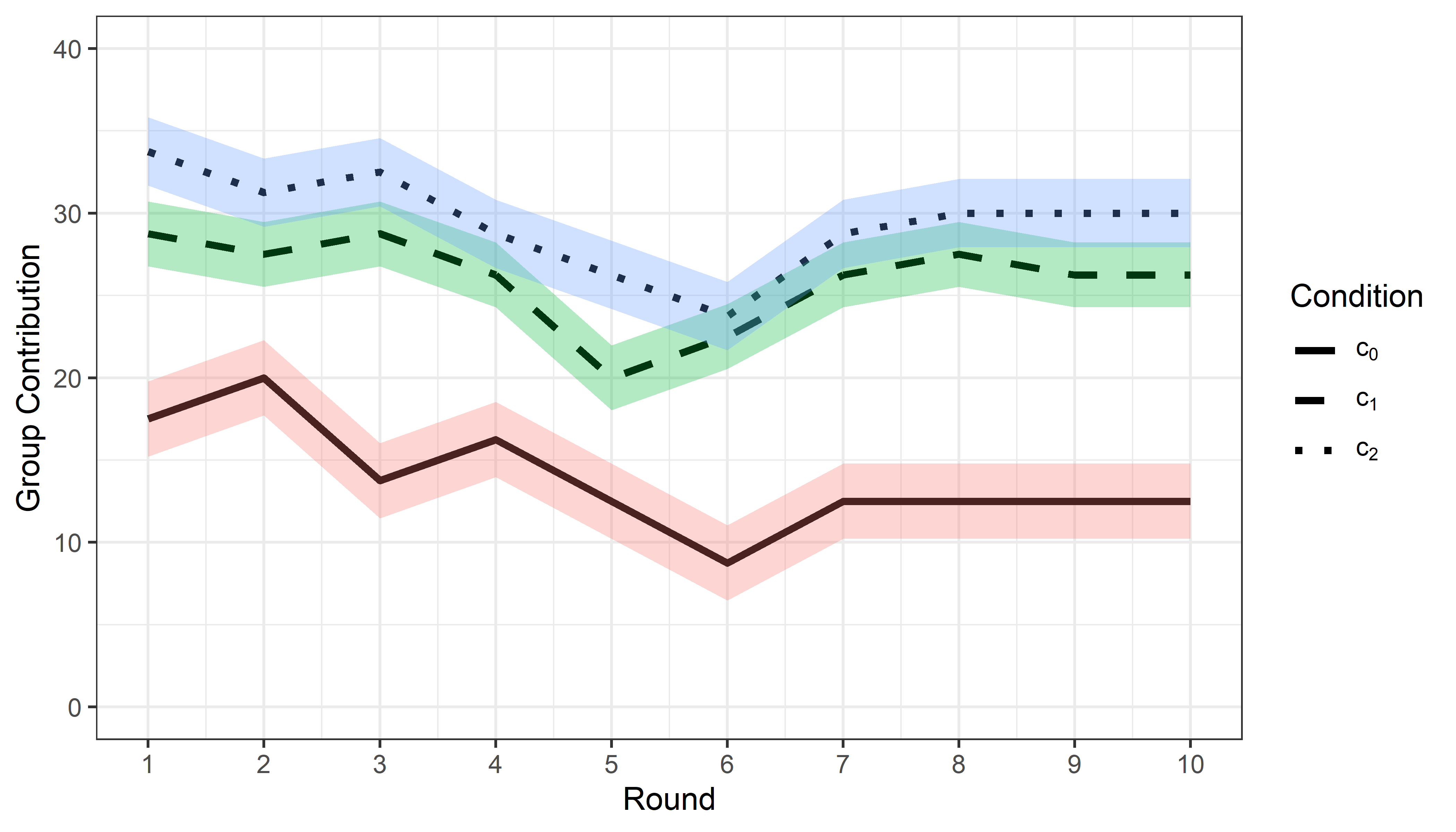}}

  \end{minipage}
  \caption{  Average group contribution  for all treatments. $c_0$ corresponds to the condition where none of the $m$ players in the sample contributed, $c_1$ to the condition where 1 out of $m$ players in the sample contributed and $c_2$ to the condition where all players in the sample contributed. The ribbon shows the 95\% confidence interval.}
  \label{fig:ttl_contributions}
\end{figure}

Figure \ref{fig:ttl_contributions} displays the aggregate average group contributions in the three treatments. The maximum possible group contribution in the experiment was 40 (when all members of the group contribute). Contributions are relatively constant across conditions, rounds and treatments, with the exception of $T_3$, where contributions fall in the first 5 rounds before stabilising for the last 5 rounds. There is no evidence of end-game effects, which is an indicator that subjects actually perceived  each round as a one-shot game. In all treatments, the same pattern is observed: contributions in condition $c_0$ are always significantly lower to contributions in $c_1$ and $c_2$.

\begin{figure}[H]\centering

\minipage{0.5\textwidth}
\subcaptionbox{$T_1$-all players}
  {\includegraphics[width=1\linewidth]{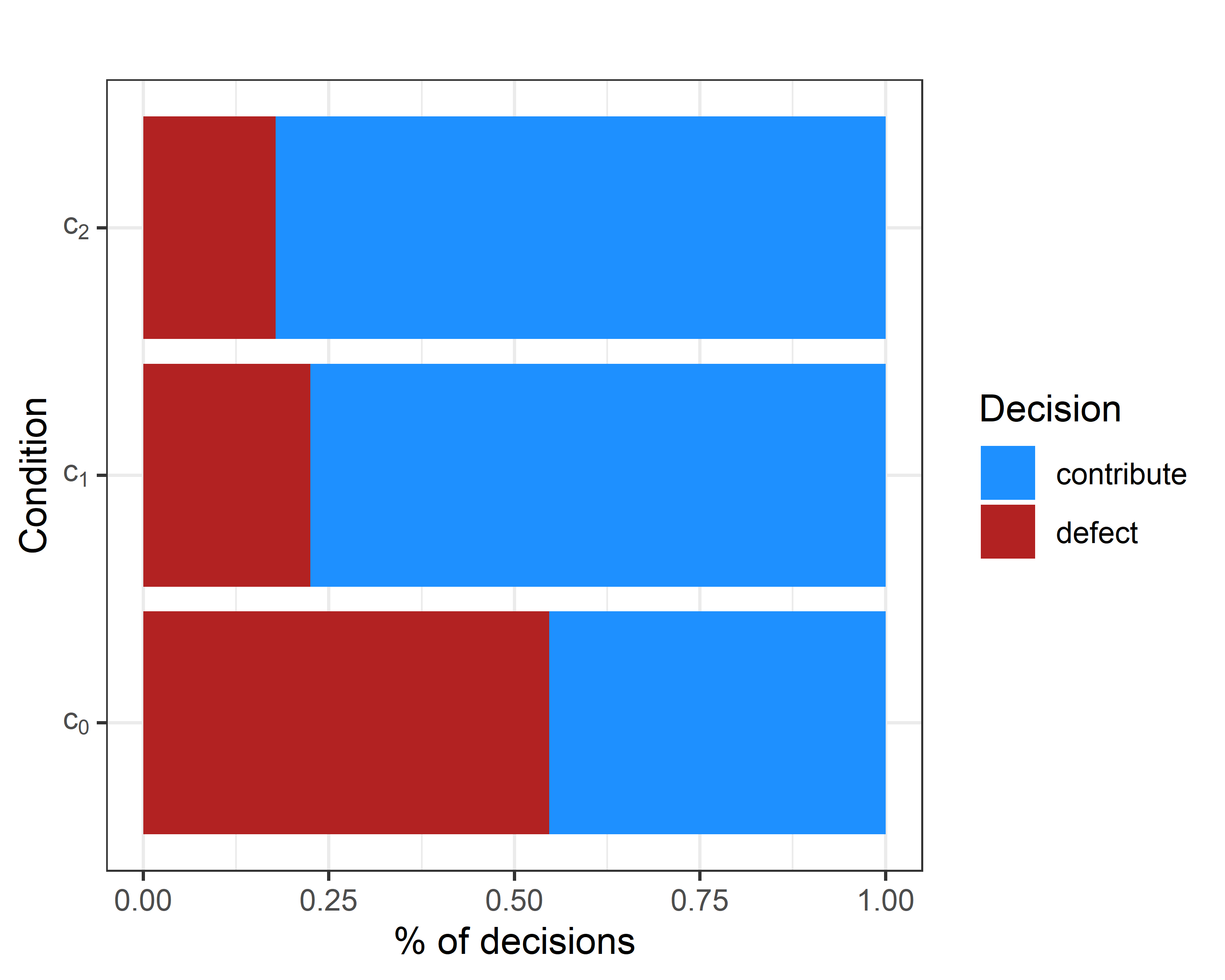}}

\endminipage\hfill
\minipage{0.5\textwidth}%
\subcaptionbox{$T_1$-players 3 \& 4}
  {\includegraphics[width=1\linewidth]{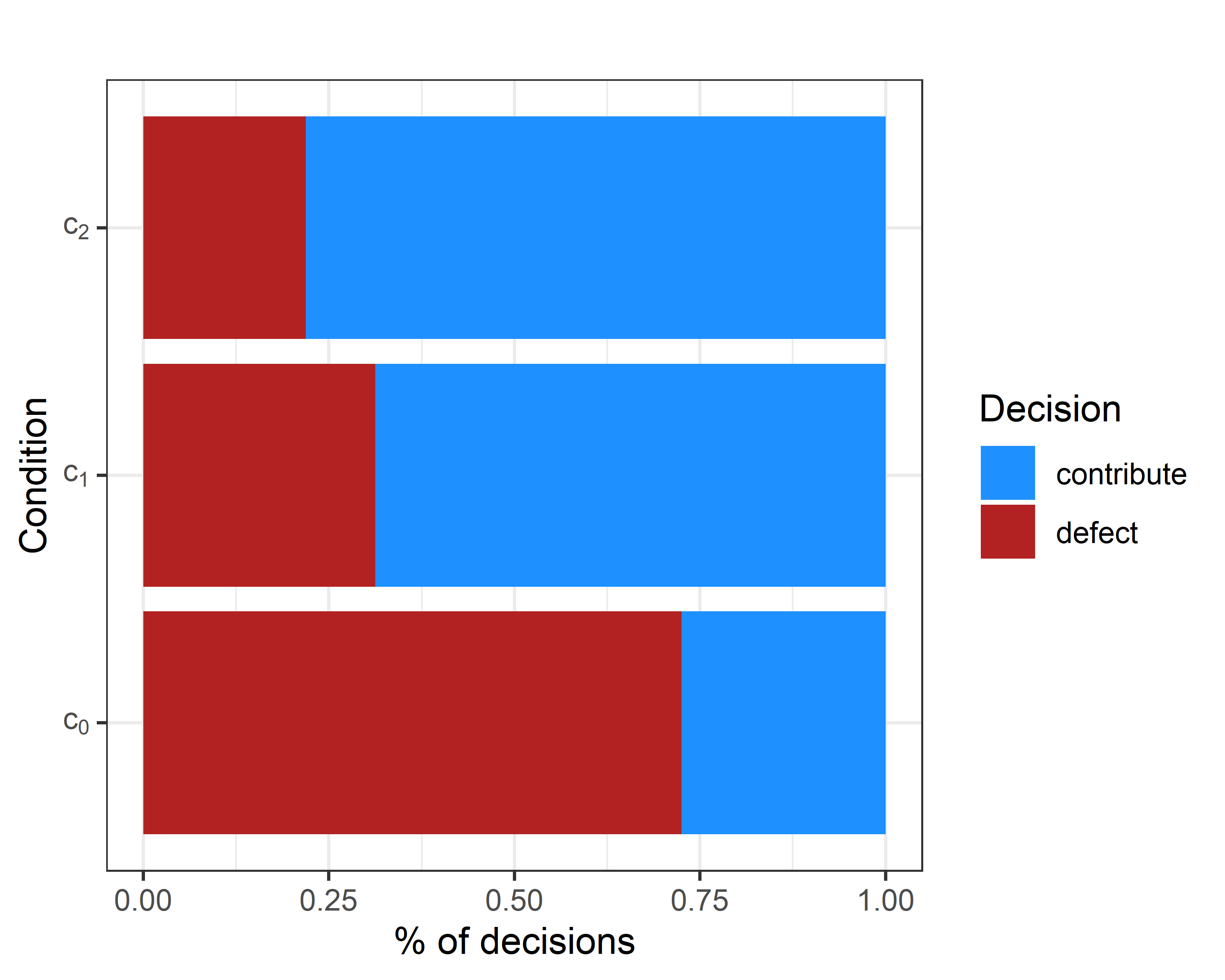}}
\endminipage\hfill

\minipage{0.5\textwidth}
\subcaptionbox{$T_2$-all players}
  {\includegraphics[width=1\linewidth]{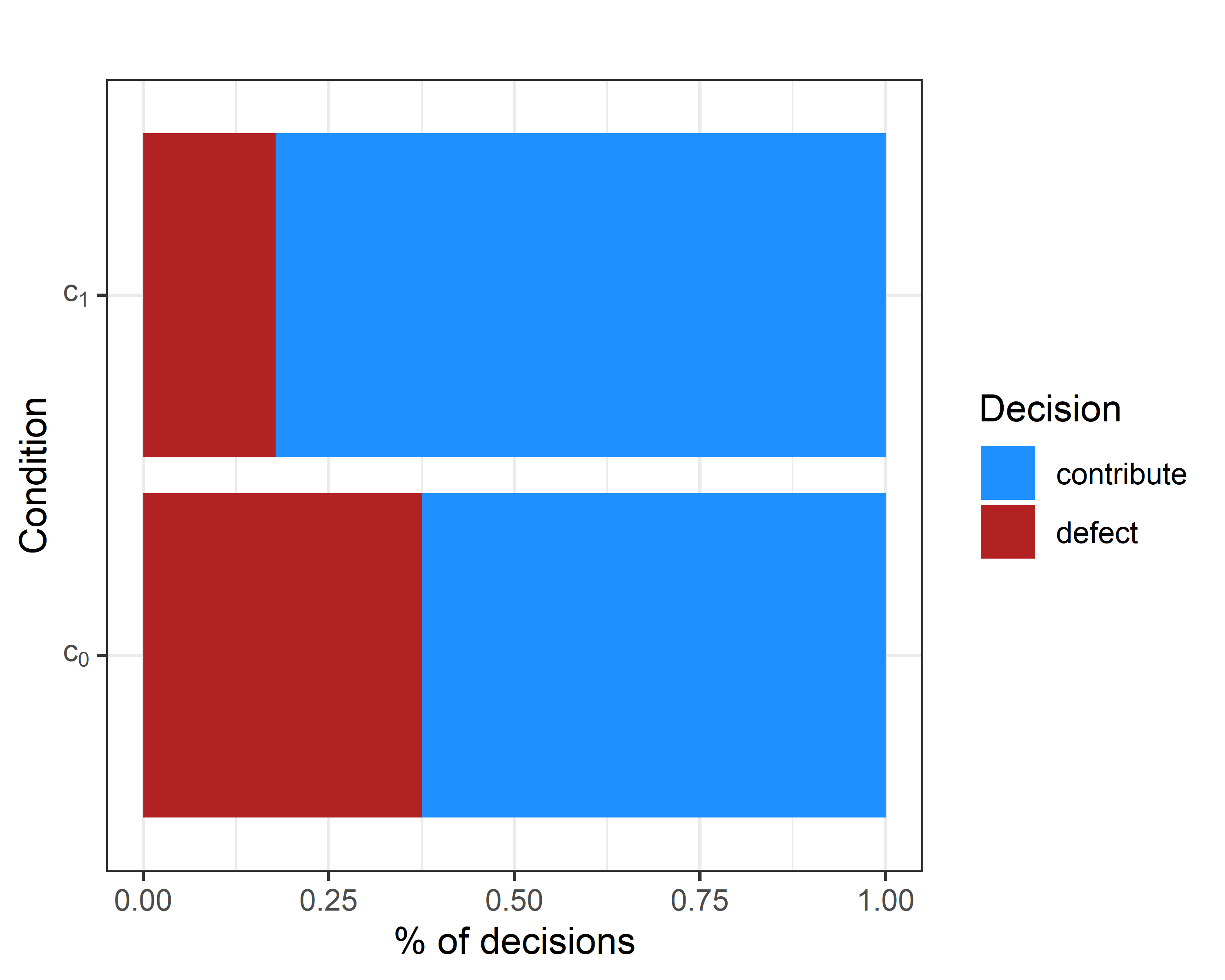}}

\endminipage\hfill
\minipage{0.5\textwidth}%
\subcaptionbox{$T_2$-players 2, 3 \& 4}
  {\includegraphics[width=1\linewidth]{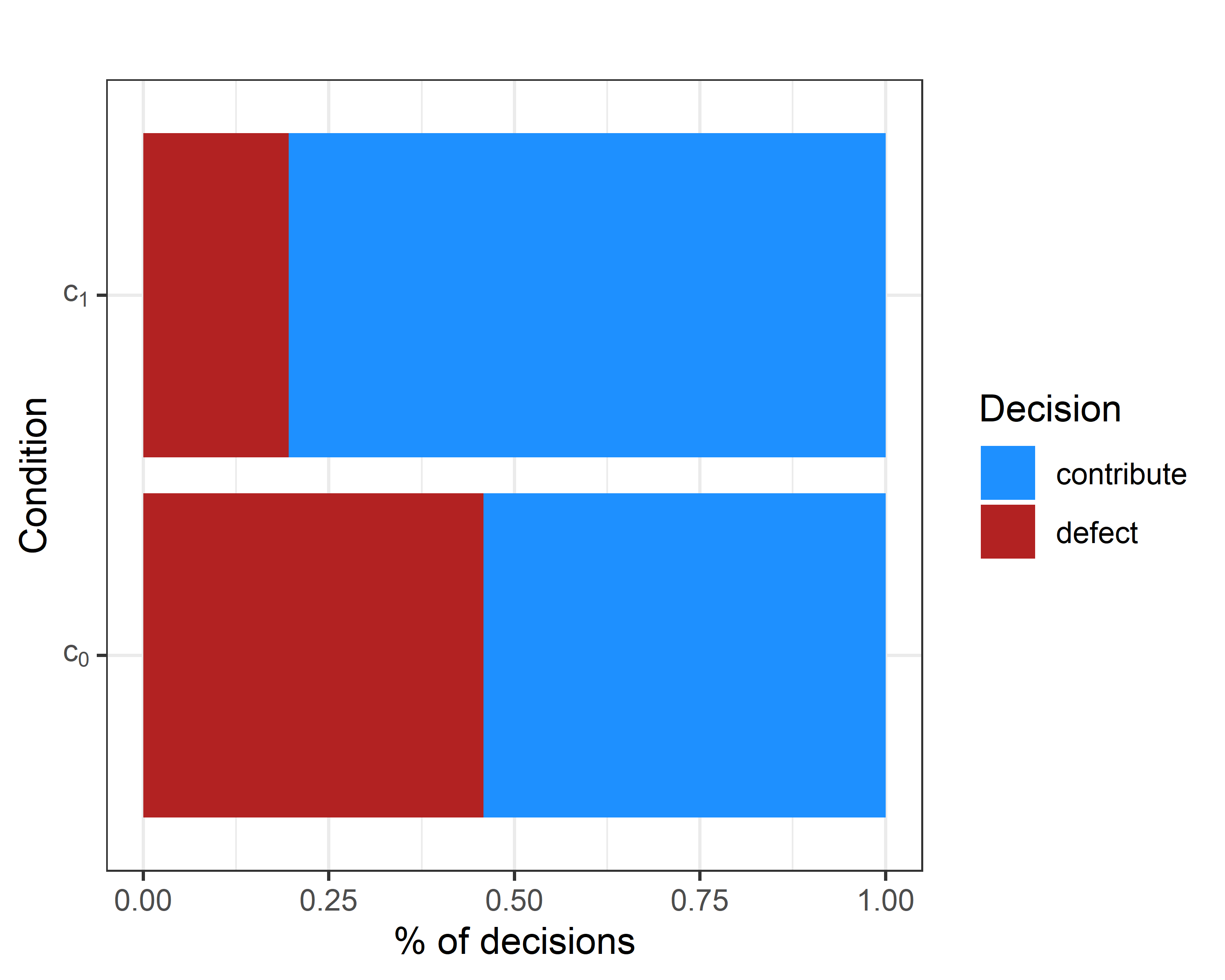}}
\endminipage \hfill

\minipage{0.5\textwidth}
\subcaptionbox{$T_3$-all players}
  {\includegraphics[width=1\linewidth]{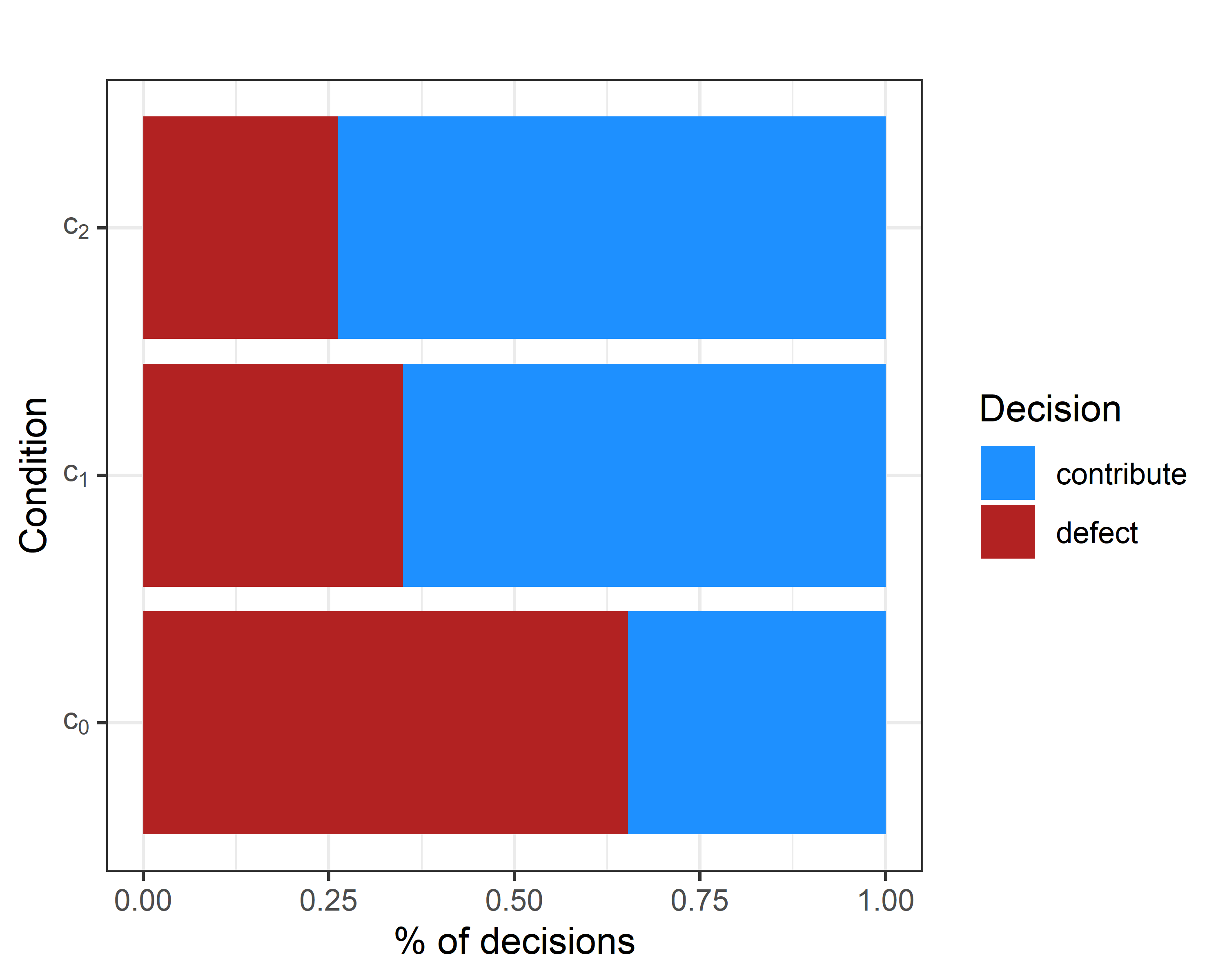}}

\endminipage\hfill
\minipage{0.5\textwidth}%
\subcaptionbox{$T_3$-player  4}
  {\includegraphics[width=1\linewidth]{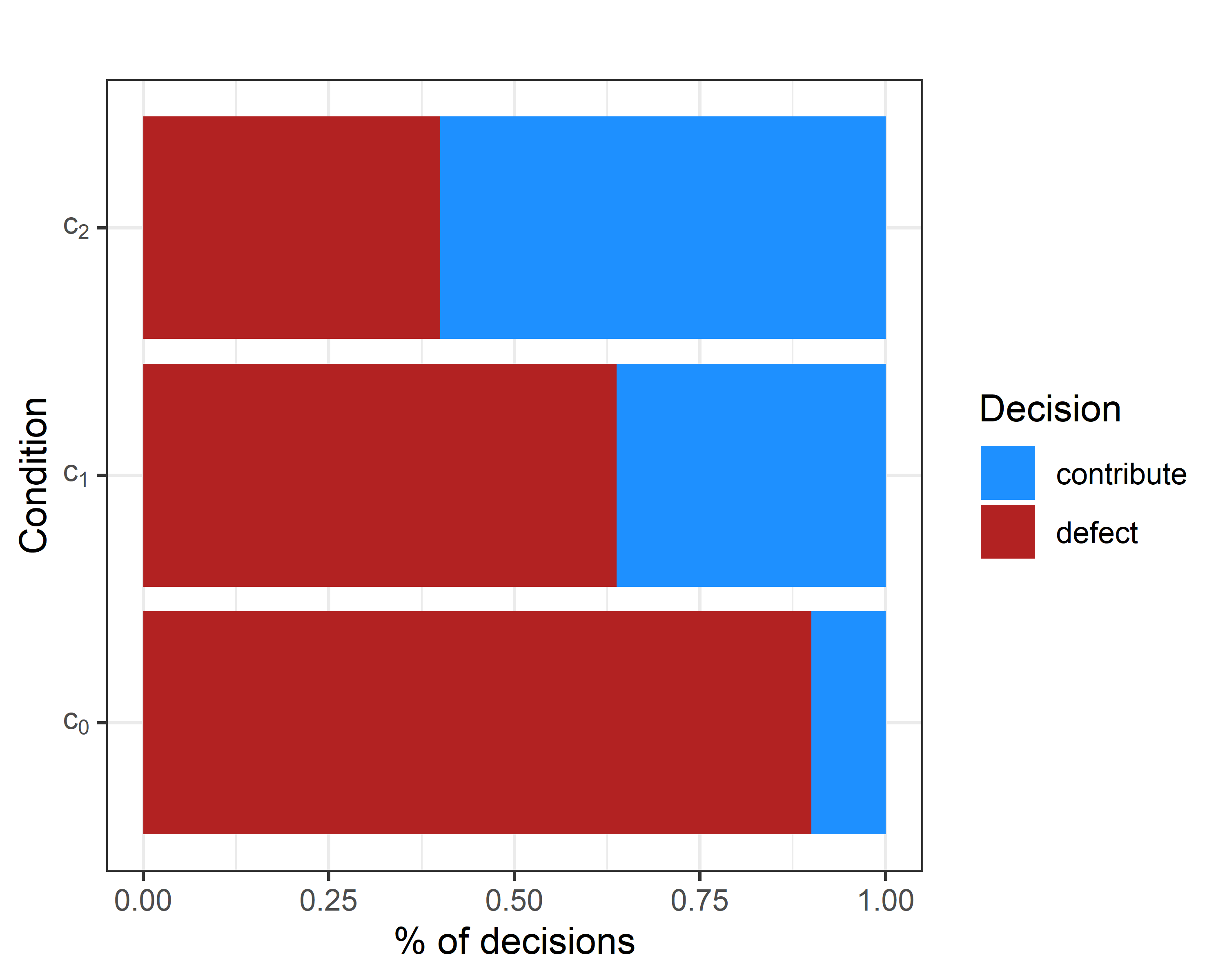}}
\endminipage \hfill

\caption{Proportion of decisions per treatment and condition. $c_0$ corresponds to the condition where none of the $m$ players in the sample contributed, $c_1$ to the condition where 1 out of $m$ players in the sample contributed and $c_2$ to the condition where all players in the sample contributed. The left panels show the decisions of all players, while the right panels only for players who face position uncertainty (for $T_1$ and $T_2$) and the last player in $T_3$.}
\label{fig:decisions}
\end{figure}

Figure \ref{fig:decisions} presents the decisions of co-operation or defect, across all treatments and conditions. The panels on the left include the choices of all the subjects, while on the right only the choices of the subjects placed at the last positions of the sequence (players 3 and 4 for $T_1$, players 2, 3 and 4 for $T_2$, and only player 4 for $T_3$). In all treatments, the  same pattern is observed. Contribution (defection) is increasing (decreasing) in the condition. The more contribution subjects observe in their sample, the more they are willing to contribute themselves. In addition, the position in the sequence has a significant effect on the decision to defect or not.\footnote{We tested this using a two sample proportion Z-test.} For instance, in condition $c_0$, players who were aware of their position, defected significantly less compared to those at the last positions. In $T_1$, only 36.9\% of decisions at positions 1 and 2 were defection, compared to 72.5\% in positions 3 and 4. Similarly, in $T_2$, only 6.3\% of the subjects who had complete information on their position defected, while those with uncertainty defected 68.8\% of the time. In $T_3$, the level of defection was relatively similar, albeit significantly different between positions, with 57.1\% (65.3\%) for players in positions 1-3 (4).

We next investigate differences within and between treatments.\footnote{Given that our dependent variable is categorical (contribute, defect) the most appropriate test is the Chi-square test for independence, when we test across different samples, while when the  data are coming from the same population, the McNemar test will be used to evaluate hypotheses about the data. Both are non-parametric tests.}   Let ${Y_j}^{c_k}$ be the proportion of contribution decisions in treatment $j$, condition $k$, with $k \in\{0,1,2\}$. In $T_1$, the prediction is that ${Y_1}^{c_0}<{Y_1}^{c_1}<{Y_1}^{c_2}$. This proportion is equal to 0.453 tokens in $c_0$, 0.775 in $c_1$ and,  0.821 in $c_2$, which is in line with the theoretical prediction. All the differences are statistically significant, based on a McNemar test, for all pairwise comparisons\footnote{Alternatively, one can use the Mann-Whitney U Test (Wilcoxon Rank-Sum Test) and compare the differences using the average contribution as a measure. We confirmed that the results are qualitatively identical.}. Both ${Y_1}^{c_1}$ and ${Y_1}^{c_2}$ are statistically different when compared to the defection sample ${Y_1}^{c_0}$ ($p<0.000$), and ${Y_1}^{c_1}$ is statistically different to ${Y_1}^{c_2}$ ($p=0.037$) at the 5\% significance level. This provides evidence in favour of Hypothesis \ref{hyp:hyp1}. Nonetheless, the finding that  ${Y_1}^{c_1}$ is different than ${Y_1}^{c_0}$ is against the theoretical prediction, where it is expected that subjects will defect in the case where they observe at least one defection.
\begin{fnd}
The contribution is higher when subjects observe full co-operation in their samples. Subjects are still contributing even if they observe defection in their samples.
\end{fnd}

For $T_2$, when the sample size is equal to 1,  the theoretical prediction is ${Y_2}^{c_0}<{Y_2}^{c_1}$. In this treatment, the proportion of contribution is 0.625 tokens in condition $c_0$, and  0.822 in $c_1$ ($p<0.000$). Next, we constraint our sample, within $c_0$, only to those subjects who observed sample of size one and defection (we drop player 1). The theory predicts that, given the rate of return used in the experiment ($r=3$), there is a mixed strategy where players contribute with probability $\gamma=0.528$ after observing a defection. In this sample, subjects contributed in 130 out of 240 observations (0.542). To test for the presence of mixed strategies, we follow \citet{rapoport92} and use a binomial test. We cannot reject the null hypothesis ($p=0.698$) $H_0$ of Hypothesis \ref{hyp:hyp2} that players play a mixed strategy. To assess the equivalence of the observed contribution rate to the mixed strategy theoretical prediction, we conducted an additional Bayesian analysis. Assuming a non-informative Beta prior, we estimated the proportion parameter $p$ through 3000 posterior draws with a thinning rate of 2. The resulting 95\% Highest Posterior Density (HPD) interval allows for a nuanced exploration of uncertainty around the estimated proportion. Notably, with an estimated HPD Interval for Proportion of (0.478, 0.604), we find robust support in favour of similarity between the observed contribution rate and the theoretical prediction.

\begin{fnd}
We cannot reject the hypothesis that subjects play a mixed strategy when they observe sample of size one and defection, providing evidence in favour of Hypothesis \ref{hyp:hyp2}.
\end{fnd}
When we include the contribution of all subjects who act as player 1, the total proportion of contribution increases from 0.542 to 0.625 tokens, in line with the theoretical prediction that the first player in the sequence will always contribute in order to motivate the subsequent players to contribute. Indeed, the proportion of contribution of the subjects who acted as player 1 is 0.875, with subjects contributing in 70 out of 80 observations, in line with previous results confirming the existence of sequence effects.

$T_3$ has identical structure to $T_1$, regarding the sample size, with the only difference being that in $T_3$, the players are aware of their position. Here, the main focus is on the contributions of player 4. The theoretical prediction, given the parameters used in the experiment, is that players placed in the last position will free-ride, and therefore, all the remaining $n-1$ players will defect as well. First, we compare  the proportion of contribution between both treatments, in condition $c_2$ (full  contribution sample of size 2). The theoretical prediction is that $Y_1^{c_2}>Y_3^{c_2}$, since the last player will always defect, and therefore, full contribution will unravel. The proportion in $T_3$ is 0.738 compared to 0.822 in $T_1$. Using a Chi-square test, we find that this difference is significant ($p=0.013$), in line with the theoretical prediction. Focusing on individual contributions at each position in the sequence,  we find a sharp decrease in the proportion of contribution for players positioned later in the sequence, namely 0.838 in position 1, 0.738 in position 2,  0.775 in position 3, and only 0.600 in the last position,  where in 52 out of 80 cases in total (65\%), subjects in the last position defected. This result is in line with \citet{figuieres2012vanishing}, in the case of full information, who find that the
average contribution declines with the position
in their sequential contribution game.
\begin{fnd}
The results align with Hypothesis \ref{hyp:hyp3} and are supportive of the prediction that full contribution unravels when subjects are aware of their position in the sequence.
\end{fnd}

\begin{figure}[h] \centering
\includegraphics[scale=1]{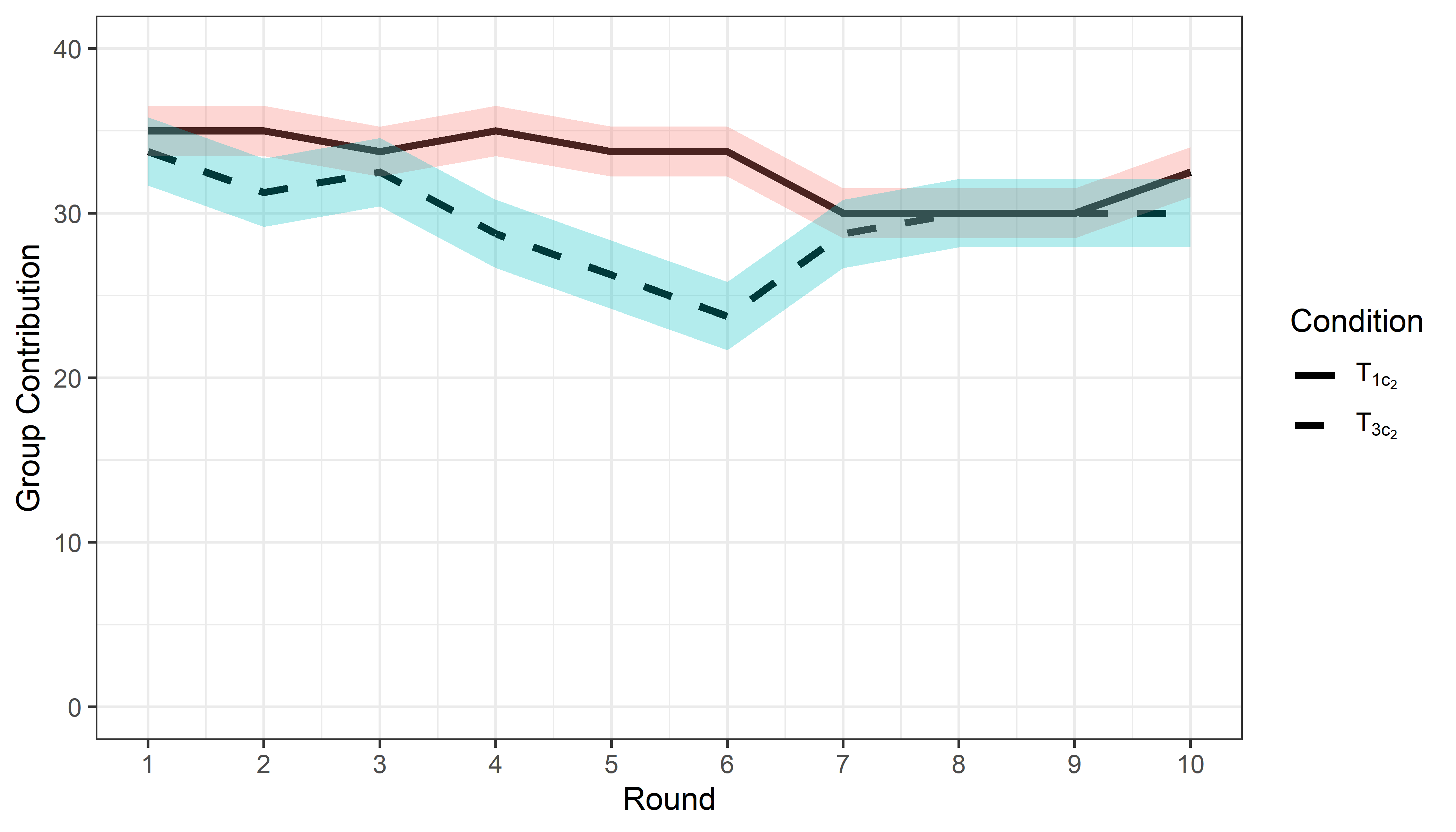}
\caption{Average group contribution in $T_1$ and $T_3$ when the provided information is that all participants in the sample contributed (condition $c_2$). The ribbon shows the 95\% confidence interval.}
\label{fig:ntools_bai}
\end{figure}

To get a better understanding on the determinants of contribution, in Appendix \ref{sec:regression} we report the results of regression analysis.
\subsection{Structural Estimation}
While the comparative statics seem to be in line with the predictions of the G\&M model, there is still a lot of unexplained heterogeneity that the model cannot account for\footnote{While  the structural analysis was not part of the hypotheses to be tested in our preregistration, in the exploratory analysis section we have mentioned that "If the theoretical predictions of the model will be rejected by our experimental data, we plan to explore whether alternative behavioural models can explain the observed data (e.g. social preferences, conditional cooperation).".}. For instance, despite the fact that the model predicts zero contribution in condition $c_0$, for all treatments, we still observe a significant number of subjects willing to contribute.  Furthermore, alternative models (types of decision makers) may be able to predict a similar behaviour pattern. In this sub-section we use structural econometric modelling to test the predictions of the G\&M model against alternative models (strategies). To achieve so, and given the discrete nature of our data, we use the \textit{Strategy Frequency Estimation Method} (SFEM), an estimation procedure introduced in \citet{dalbo11} and \cite{fudenberg12} and has been extensively used since.\footnote{A non-exhaustive list of studies includes \citet{bigoni2015time}, \citet{breitmoser2015}, \citet{frechette2017infinitely}, \citet{dalbo2019}, using this method to estimate the frequency of individual strategies in co-operation games. \citet{bardsley2007experimetrics} have proposed a similar methodology in the context of continuous public goods games estimating a mixture model on a number of pre-determined types. \citet[p. 3930]{dalbo2019} provide an overview of alternative methods to estimate the use of strategies, discussing the various identification issues that characterise them.} The SFEM method first specifies a set of candidate strategies and then estimates their frequencies in a finite-mixture model, allowing for the possibility of implementation errors. Formally, the SFEM results provide two outputs, $\pi$ and $\beta$, both at the population level: $\pi$ is a probability distribution over the set of strategies, and $\beta$ is the probability that the choice corresponds to what the strategy prescribes (more details below).

 Following the terminology of types adopted in the type-classification literature in public goods games (see \citealt{fischbacher2001people}; \citealt{bardsley2007experimetrics}; \citealt{thoni2018conditional}; \citealt{katuscak2023drives}; \citealt{preget2016}) we define and explore the existence of the following types in our data:
\begin{itemize}
\item The G\&M type, who behaves as presented in section \ref{sec:theory}.
\item The free-rider, who never contributes no matter the position.
\item The altruist, who always contributes no matter the position.
\item The conditional co-operator, who always contributes if she is in position 1 and contributes if at least one other person in the sample has contributed when she is in positions 2-4.
\end{itemize}

While the free-rider and the altruist type definition is straightforward, the conditional co-operator type deserves some further discussion. This type contributes when at least one other subject in the sample has contributed. The main difference between this type and the G\&M type is that the former  always contributes when only one of the previous two players have contributed (in the case of sample size of 2, where the G\&M type defects), while in the case of sample size of 1, the conditional co-operator defects, if the previous player defected, and the G\&M type contributes with probability $\gamma$. Finally, since the G\&M model predicts that contribution will unravel when players are aware of their position in the sequence, the G\&M's behaviour coincides to that of a free-rider in $T_3$.   Table \ref{tab:strategies} lists the strategy profile for each type.

\begin{table}[h]
\centering
\begin{tabular}{@{}lc|cc|ccc|ccc@{}}
\toprule
\multicolumn{1}{l}{Position} & $P_1$ & \multicolumn{2}{c|}{$P_2$} & \multicolumn{3}{c|}{$P_3$} & \multicolumn{3}{c}{$P_4$} \\ \midrule
\multicolumn{1}{l}{Type} & $c_0$ & $c_0$ & $c_1$ & $c_0$ & $c_1$ & $c_2$ & $c_0$ & $c_1$ & $c_2$ \\
G\&M ($T_1$) & 1 & 0 & 1 & 0 & 0 & 1 & 0 & 0 & 1 \\
G\&M ($T_2$) & 1 & $\gamma$ & 1 & $\gamma$ & 1 & - & $\gamma$ & 1 & - \\
G\&M ($T_3$) & 0 & 0 & 0 & 0 & 0 & 0 & 0 & 0 & 0 \\
Free rider & 0 & 0 & 0 & 0 & 0 & 0 & 0 & 0 & 0 \\
Altruist & 1 & 1 & 1 & 1 & 1 & 1 & 1 & 1 & 1 \\
Conditional co-operator ($T_1$ \& $T_3$) & 1 & 0 & 1 & 0 & 1 & 1 & 0 & 1 & 1 \\
Conditional co-operator ($T_2$) & 1 & 0 & 1 & 0 & 1 & - & 0 & 1 & - \\ \bottomrule
\end{tabular}
\caption{Strategy profile for all the types. $P_t$ indicates the player in position $t$ and $c_k$ the condition (information on past contributions) with $k \in \{0,1,2\}$. $\gamma$ is the mix probability of contributing.}
\label{tab:strategies}
\end{table}

The SFEM assumes a pre-determined set of strategies $\mathcal{K}$ and that a subject $i$ chooses strategy $k \in \mathcal{K}$ with probability $\pi_k$. In each round the subject plays according to the chosen strategy with probability $\beta \in (1/2,1)$ and makes a mistake with probability $(1-\beta)$. The likelihood that the observed data were generated by strategy $k$ is given by:
\[P(s^k)=\prod_{R} \beta^{I_{iR}^k}(1-\beta)^{1-I_{iR}^k}\]
with $R\in\{1, \cdots, 10\}$ representing a particular round in the experiment, and $I_{iR}^k$ being an indicator function taking the value 1 when the predicted choice of strategy $k$ in round $R$ agrees with the subject's $i$ actual choice, otherwise it takes the value 0. Objective is to find optimal values for $\beta$ and $\pi$ that maximise the overall likelihood across all subjects $I$ and set of types (strategies) $\mathcal{K}$:
\begin{equation}
\max_{\beta,\pi}\sum_I \ln \left(\sum_K\pi^kP(s^k)\right)
\end{equation}
We assume that tremble probabilities are partitioned by treatment. Therefore, there are 4 parameters to estimate for each of the treatments, the tremble $\beta$, and the mixing probabilities $ \pi_{gm}, \pi_{alt}$ and $\pi_{coop}$ for the G\&M, the altruist and the conditional co-operator type respectively (the mixing probability $\pi_{free}$ for the free-rider is simply the residual probability), giving in total 12 parameters.\footnote{ \citet{fudenberg12} recognises that this specification assumes that all subjects are identical in terms of the probability distribution over strategies and over errors, while \citet{bland20} shows how allowing the distribution of trembles to vary, leads to more robust estimates. We also estimated a model where we allow for heterogeneous trembles (i.e. assume that each subject's tremble probability $\beta_i$ is drawn from a treatment-specific population distribution with mean $\mu$ and standard deviation $\sigma$). The results are quantitatively similar as the estimates of the standard deviations were very close to zero, with the only difference that the estimate of the G\&M proportion was slightly higher in $T_1$. The results are available upon request.
} We estimate the model using Maximum Likelihood Estimation techniques.\footnote{For the estimation we use a general nonlinear augmented Lagrange multiplier optimisation routine that allows for random initialisation of the starting parameters as well as multiple restarts of the solver, to avoid local maxima. The estimation was conducted using the \emph{R} programming language for statistical computing (The \emph{R} Manuals, version 4.3.1. Available at: http://www.r-project.org/). The estimation codes are available in the replication depository.}

Table \ref{tab:estimates} reports the estimates. The results are ordered by treatment with $\beta_j$ being the error probability for treatment $j$ and $\pi^j_k$ the mixing probability for type $k$ in treatment $j$\footnote{To explore the robustness of the SFEM methodology we conducted an extensive Monte Carlo simulation. For details see Appendix \ref{sec:simulation}.}.

\begin{table}[H]
\centering
\begin{tabular}{@{}lccccc@{}}
\toprule
\multicolumn{1}{c}{Parameter} & Estimate & s.e. & p-value & lower & upper \\ \midrule
$\beta_1$ & 0.866 & 0.013 & 0.000 & 0.838 & 0.894 \\
$\pi^1_{gm}$ & 0.176 & 0.075 & 0.033 & 0.016 & 0.336 \\
$\pi^1_{alt}$ & 0.319 & 0.088 & 0.002 & 0.132 & 0.506 \\
$\pi^1_{coop}$ & 0.442 & 0.095 & 0.000 & 0.240 & 0.644 \\
$\pi^1_{free}$ & 0.063 & 0.149 & 0.679 & -0.256 & 0.382 \\
$\beta_2$ & 0.893 & 0.015 & 0.000 & 0.861 & 0.925 \\
$\pi^2_{gm}$ & 0.258 & 0.099 & 0.019 & 0.048 & 0.468 \\
$\pi^2_{alt}$ & 0.406 & 0.097 & 0.001 & 0.199 & 0.613 \\
$\pi^2_{coop}$ & 0.242 & 0.085 & 0.012 & 0.062 & 0.422 \\
$\pi^2_{free}$ & 0.094 & 0.162 & 0.571 & -0.252 & 0.440 \\
$\beta_3$ & 0.819 & 0.015 & 0.000 & 0.788 & 0.850 \\
$\pi^3_{gm}$ & 0.258 & 0.079 & 0.005 & 0.083 & 0.427 \\
$\pi^3_{alt}$ & 0.072 & 0.057 & 0.225 & -0.050 & 0.194 \\
$\pi^3_{coop}$ & 0.669 & 0.097 & 0.000 & 0.461 & 0.877 \\\hline
Log Likelihood & 928.830    &  &    &  &  \\
Num. pars & 11    &  &    &  &  \\
Num. Obs. & 2000    &  &    &  &  \\\bottomrule
\end{tabular}
\caption{The Table reports the SFEM estimates. $\beta_j$ is the noise parameter for treatment $j$. $\pi^j_{k}$ is the mixing probability (proportion of subjects) of strategy $k$ in treatment $j$. \(gm\) stands for the G\&M type, \(alt\) for the altruist, \(coop\) for the conditional co-operator, and  \(free\) for the free rider. }
\label{tab:estimates}
\end{table}
The first thing to notice is the high value of $\beta$ for all treatments, ranging between 0.819 and 0.893. This parameter is always significant and different from 0.5 ($\beta$ values close to 0.5 indicate random behaviour, while values close to unity indicate almost deterministic behaviour). There is also little variation between treatments regarding the error probability. As a general overview of the results, the fraction of behaviour consistent with the G\&M type ranges between 17.6\% and 25.8\% across all treatments\footnote{This range increases to 20\% and 26.1\%, when a heterogeneous error probability is assumed.}, the fraction of the altruists between 8.7\% and 40.6\%, that of conditional co-operators between 24.2\% and 66.9\% and a very small and insignificant proportion of subjects is classified as free riders. In particular, when there is position uncertainty ($T_1$ and $T_2$) the vast majority of the subjects are classified either as altruists or conditional co-operators 76.1\% (64.8\%) in $T_1$ ($T_2$). This is in sharp contrast with the G\&M model prediction, that subjects will defect if they observe at least one defection in their sample. While the behaviour of a conditional co-operator is straightforward (i.e. reciprocate to the existing contribution), the behaviour of the altruist is worth further exploration. On top of altruistic motives, a potential explanation could be that subjects were trying to signal to members of the group, placed later in the sequence, and motivate them to contribute.\footnote{This is in line with \citet{cartwright2010} who use a sequential public goods game with exogenous ordering and show that agents early enough in the sequence who believe imitation to be sufficiently likely, would want to contribute.}  Support for the latter is given by the estimated proportion of altruists when the position is known ($T_3$), where it drops to virtually zero (7.2\% and insignificant), where due to the Treatment's characteristics, contributing unconditionally does not seem to be an appealing  strategy, as is the case when only the action of a past player is revealed (Treatment 2). If subjects expect the last player in the sequence to defect, they lack the motivation to unconditionally  contribute. On the contrary, the proportion of  conditional co-operators rises to 66.9\% when there is position certainty. The G\&M model fits well for around 1/4 of the experimental population (25.8\% in both $T_2$ and $T_3$), while it accounts for 17.6\% of the subjects, when the sample is equal to 2 and there is position uncertainty ($T_1$). A potential explanation of this drop could be linked to the strict prediction of the model, that one should ignore any contributions in the sample and defect instead, if there is at least one defection.  Finally, the proportion of free riders in the experimental population is virtually zero in $T_1$ and $T_2$, as the fraction of free riders is estimated to be very low and is always insignificant. This result is in line with  \citet{dalbo2019} who find that the strategies that represent less than 10 per cent of the data
are rarely identified as statistically significant. Nevertheless, the information structure of $T_3$ appears to motivate  free-riding behaviour, around 1/4 of our experimental population is classified as $G\&M$ or free-riders.
 In short, the main conclusion from this type classification in the simultaneous public goods games literature is that conditional co-operation is the predominant pattern, free-riding is frequent, while unconditional co-operation is very rare. In a sequential discrete public goods experiment involving position uncertainty or position certainty with a partial lack of information on past contributions, we demonstrate that the majority of subjects exhibit altruistic or conditional cooperating behaviour. Approximately 25\% of subjects behave as G\&M-type individuals, while free-riding is found to be very rare, unless the position in the sequence is known.
\begin{fnd}
The majority of the subjects behave in an altruistic or conditional co-operating way, around 25\% of the subjects as G\&M type, and free-riding is very rare.
\end{fnd}
Finally, from a mechanism design point view, we explore which information structure of this kind of contribution mechanism can maximise the provision of the public good. The average contribution across all information conditions is 6.833, 7.234 and  5.781 tokens, for $T_1$, $T_2$ and $T_3$ respectively, in line with  the previous findings that position certainty leads to higher percentage of defection. The difference between $T_1$ and $T_3$, and $T_2$ and $T_3$ is always statistically significant (Wilcoxon rank sum test \(p<0.000\)), while the difference between $T_1$ and $T_2$ is significant only at the 10\% level $(p=0.087)$.	
\begin{fnd}
The average provision of the public good is maximised when agents are not aware of their position in the sequence, and observe the action of their immediate one predecessor. 
\end{fnd}
\section{Conclusion}\label{conclusion}
In this study, we present the results of an economic experiment designed to
test the theoretical predictions of \citet{gallice2019co}. In the framework of a sequential public goods game, the experiment contributes to the understanding of contribution decisions when the agents receive partial information regarding their predecessors' past actions, and are uncertain about their position in the sequence.
In a pre-registered economic experiment, we conducted three treatments, varying the
amount of information about past actions that a subject can observe, as well as their positional awareness. Using rigorous structural econometric analysis, we find that approximately
25\% of the subjects behave according to the theoretical predictions of \citet{gallice2019co}. Allowing for the presence of alternative behavioural types among the remaining subjects, we find that the majority are classified as conditional co-operators, some are altruists, and very few behave in a free-riding way.

The rationale for choosing these types is twofold. Firstly, they are commonly assumed in the literature, and prior research across various settings and games has provided empirical support. Secondly, as they are parameter-free, incorporating them into our SFEM analysis within our simple discrete public goods experimental design is straightforward. Of course, there might be alternative models that capture other psychological processes and can generate similar decision patterns (or explain the remaining heterogeneity). For instance, the presence of \textit{dispositional trust}, people’s general propensity to trust others, is a potential candidate model (see \citealt{rotter1967}). Research in social psychology and organisational behaviour (see among others \citealt{dawes1980}; \citealt{yamagishi1986}; \citealt{bianchi2012}) has provided ample evidence that people who are more trusting are likely to have more positive fairness perceptions, while evidence from neuroscience using fMRI data shows that dispositional trust positively affects prosocial decisions to cooperate \citep{emonds14}. 

Trust is usually modelled as a subjective probabilistic belief about the trustworthiness of others (\citealt{bhattacharya1998formal}; \citealt{eckel2004trust}; \citealt{fetchenhauer2009trust}), and the estimation and testing of such models call for the use of auxiliary measures of trust, either incentivised (i.e., trust game, elicitation of beliefs) or non-incentivised self-reported measures of trust (e.g., \citealt{glaeser2000measuring}). Research in experimental economics has explored the role of dispositional trust, either by correlating measures of trust to levels of cooperation as in \citet{gaechter2004trust}, or by incorporating measures of trust into structural models of reciprocity as in \citet{guzman2020game}\footnote{Nevertheless, how to efficiently measure trust remains an open question. \citet{hancock2023how} raise a number of methodological issues, such as the temporal stability of trust depending on whether it was measured before, during, or after an interaction.}. As our main objective was to test the theoretical prediction of the G\&M model, in an effort to keep the experimental protocol as simple as possible, we did not include any auxiliary trust elicitation measures. Of course, richer experimental protocols (e.g., continuous contributions, trust surveys, elicitation of beliefs), would be able to provide further insights into the role of trust in sequential public goods games.

We also find that on average provision of public goods is maximised when agents are not aware of their position and agents only observe a partial sample of their immediate predecessor. This might have some implications for designing a contribution mechanism with an information structure akin to G\&M, in order to improve the efficiency of public good provision. For instance, in the crowdfunding campaign to improve conservation example, discussed in the introduction, it could be beneficial to set up a website mimicking the information structure proposed by G\&M. If potential contributors are not aware of where they are in the sequence, and only partial history of the immediate predecessor is revealed to them before they make contributions, then it would be possible to maximise the voluntary provision of the public good.

Our study's exploration into the role of positional uncertainty in sequential public goods games highlights its impact on cooperative behaviour (Findings 1, 2, 5). These findings align with broader research on social dilemmas, particularly underlining the relevance of the Constrained-Egoism/ Greed-Efficiency-Fairness Hypothesis. This framework helps explain why, in the face of positional uncertainty, individuals may balance self-interest with fairness, resulting in increased cooperation (Findings 1 and 2) and optimal public good provision (Finding 5). Such behaviour contrasts with responses to resource size uncertainty, where the egoism justification hypothesis might prevail, leading to non-cooperative actions. Our results, including the unravelling of full contribution with position certainty (Finding 3) and the prevalence of altruistic or conditional cooperation (Finding 5), underscore how different uncertainties distinctly influence cooperative dynamics. This psychological perspective sheds light on the complex decision-making processes in cooperative settings, contrasting with scenarios involving resource size uncertainty.

In considering the varied responses to uncertainties, psychological theories offer insights. Positional or group size uncertainties, as studied in our research and in \citet{au2003effects}, individuals may adopt a conservative strategy due to the ambiguity in assessing their impact or the group's collective action. This leads to a reliance on fairness heuristics or risk aversion, favouring cooperative choices. Conversely, resource size uncertainty as in \citet{rapoport1993sequential}, often triggers the egoism justification hypothesis, where individuals justify prioritising self-interest due to resource unpredictability. Understanding these psychological responses to different types of uncertainties is crucial in deciphering the observed behaviours in social dilemma studies.

While there is extensive empirical literature studying sequential public goods games with perfect information, there is a lack of studies looking at cases with imperfect, incomplete or no information at all. This study is a first step towards that direction. Our experiment focuses only on a small part of the experimental space (i.e., small groups, all or none contributions, limited information about previous actions). Fruitful paths for future research involve expanding the scope of our experiment along various dimensions. Firstly, one could manipulate the return from contributions coefficient to systematically investigate its impact on participant behaviour. Secondly, while our current design focused on all-or-none contributions, an extension to continuous contributions, coupled with the elicitation of beliefs and trust measures, would enrich the dataset. This expansion would not only enable a more detailed exploration of trust dynamics but also facilitate the estimation of more elaborate models of other-regarding preferences. Lastly, considering the current narrow focus on student subject pools and relatively low stakes, future research could explore the external validity of the mechanism. Extending the framework to non-student populations and introducing higher stakes would allow for a comprehensive examination of the generalisability of our findings. This extension is particularly crucial to assess the potential real-life applications and practical implications of the observed behavioural patterns.

\bibliography{references_PG}{}
\bibliographystyle{dcu}

\newpage
\begin{appendices}
\setcounter{table}{0}
\renewcommand{\thetable}{A\arabic{table}} 

\section{Simulation}\label{sec:simulation}
In this appendix, we use simulations to explore the robustness of the SFEM methodology employed in this paper. Following \citet[Appendix E]{dalbo2019}, we conduct an extensive Monte Carlo simulation to study the SFEM's ability to identify the prevalence of different strategies in the data and, consequently, the existence of various types in the subject pool. We assess the methodology's robustness, subject to our sample size, by successfully recovering the assumed parameters of our structural model.

We simulate data and assume conditions that resemble our lab implementation. For a given treatment, we simulate data from 32 subjects. Each subject is allocated a type based on predetermined proportions $\pi_k$ with $k\in\{gm,alt,coop, free\}$ for the G\&M, the altruist, the conditional co-operator, and the free-rider, respectively. In every round, subjects are matched to groups of 4 and are allocated to the sequence in a random order. Choices are simulated based on the experimental design. That is, each subject provides their conditional choice, dependent on their type and their position in the sequence. The process is repeated for 10 rounds, where subjects are matched to a new 4-member group. To take stochasticity in choice into consideration, we assume that subjects are choosing what their type dictates with probability $\beta \in (1/2,1)$, and with probability $1-\beta$, they make a mistake.

We assume three levels of noise: high ($\beta=0.60$), medium ($\beta=0.75$), and low ($\beta=0.90$). The frequency of types in the population of 32 subjects is set to 6 (0.188), 9 (0.281), 12 (0.375), and 5 (0.156) subjects for the G\&M, the altruist, the conditional co-operator, and the free-rider respectively, in $T_1$ and $T_2$. As the G\&M and the free-rider predict the same choice pattern in $T_3$, the frequencies are set to 7 (0.219), 11 (0.344), and 14 (0.438) subjects in $T_3$ for the G\&M, the altruist, and the conditional co-operator\footnote{We explored numerous frequency combinations, and our findings consistently affirm that qualitative results remain unchanged, regardless of the assumed frequency combination}. Based on the value of the parameter $\beta$ and the frequencies of the types, we generate treatment-level simulated datasets. Each dataset comprises the choices of 32 subjects over 10 rounds. Subsequently, the dataset is used as input for our estimation code to recover the value of $\beta$ and the estimated frequency of types. We repeat this process for a total of 100 Monte Carlo simulations.

Tables \ref{tab:sim1} to \ref{tab:sim3} report the results of the simulation for all treatments across each of the three levels of noise. In particular, the tables present the actual values of the parameters used in the simulation, the mean of 100 Monte Carlo simulations for each parameter, and their respective standard deviations.

\begin{table}[h]
\centering
\begin{tabular}{@{}llccccc@{}}
\toprule
 &  & $\beta$ & $\pi_{gm}$ & $\pi_{alt}$ & $\pi_{coop}$ & $\pi_{free}$ \\ \midrule
\multicolumn{1}{c}{\multirow{3}{*}{$T_1$}} & True & 0.600 & 0.188 & 0.281 & 0.375 & 0.156 \\
\multicolumn{1}{c}{} & Mean Estimate & 0.601 & 0.257 & 0.286 & 0.312 & 0.145 \\
\multicolumn{1}{c}{} & s.d. & 0.027 & 0.184 & 0.145 & 0.196 & 0.107 \\
 &  &  &  &  &  &  \\
\multirow{3}{*}{$T_2$} & True & 0.600 & 0.188 & 0.281 & 0.375 & 0.156 \\
 & Mean Estimate & 0.599 & 0.217 & 0.263 & 0.369 & 0.151 \\
 & s.d. & 0.028 & 0.209 & 0.172 & 0.207 & 0.126 \\
 &  &  &  &  &  &  \\
\multirow{3}{*}{$T_3$} & True & 0.600 & 0.219 & 0.344 & 0.438 & - \\
 & Mean Estimate & 0.596 & 0.195 & 0.342 & 0.464 & - \\
 & s.d. & 0.025 & 0.122 & 0.184 & 0.224 & - \\
 &  &  &  &  &  &  \\ \bottomrule
\end{tabular}
\caption{The Table reports the results from the SFEM simulation for the high noise ($\beta=0.60$). $\pi^{k}$ is the mixing probability (proportion of subjects) of strategy $k$. \(gm\) stands for the G\&M type, \(alt\) for the altruist, \(coop\) for the conditional co-operator, and  \(free\) for the free rider. The frequency of types is set to 6 (0.188), 9 (0.281), 12 (0.375) and 5 (0.156) subjects for each type respectively, in $T_1$ and $T_2$, and 7 (0.219), 11 (0.344) and  14 (0.438) subjects in $T_3$ for the G\&M  the altruist, and the conditional co-operator.}
\label{tab:sim1}
\end{table}

\begin{table}[h]
\centering
\begin{tabular}{@{}llccccc@{}}
\toprule
 &  & $\beta$ & $\pi_{gm}$ & $\pi_{alt}$ & $\pi_{coop}$ & $\pi_{free}$ \\ \midrule
\multirow{3}{*}{$T_1$} & True & 0.750 & 0.188 & 0.281 & 0.375 & 0.156 \\
 & Mean Estimate & 0.746 & 0.188 & 0.281 & 0.372 & 0.160 \\
 & s.d. & 0.016 & 0.060 & 0.038 & 0.067 & 0.023 \\
 &  &  &  &  &  &  \\
\multirow{3}{*}{$T_2$} & True & 0.750 & 0.188 & 0.281 & 0.375 & 0.156 \\
 & Mean Estimate & 0.747 & 0.174 & 0.289 & 0.383 & 0.154 \\
 & s.d. & 0.020 & 0.146 & 0.085 & 0.099 & 0.031 \\
 &  &  &  &  &  &  \\
\multirow{3}{*}{$T_3$} & True & 0.750 & 0.219 & 0.344 & 0.438 & - \\
 & Mean Estimate & 0.746 & 0.215 & 0.343 & 0.442 & - \\
 & s.d. & 0.016 & 0.017 & 0.040 & 0.042 & - \\
 &  &  &  &  &  &  \\ \bottomrule
\end{tabular}
\caption{The Table reports the results from the SFEM simulation for the medium noise ($\beta=0.75$). $\pi^{k}$ is the mixing probability (proportion of subjects) of strategy $k$. \(gm\) stands for the G\&M type, \(alt\) for the altruist, \(coop\) for the conditional co-operator, and  \(free\) for the free rider. The frequency of types is set to 6 (0.188), 9 (0.281), 12 (0.375) and 5 (0.156) subjects for each type respectively, in $T_1$ and $T_2$, and 7 (0.219), 11 (0.344) and  14 (0.438) subjects in $T_3$ for the G\&M  the altruist, and the conditional co-operator.} 
\label{tab:sim2}
\end{table}

\begin{table}[h]
\centering
\begin{tabular}{@{}llccccc@{}}
\toprule
 &  & $\beta$ & $\pi_{gm}$ & $\pi_{alt}$ & $\pi_{coop}$ & $\pi_{free}$ \\ \midrule
\multirow{3}{*}{$T_1$} & True & 0.900 & 0.188 & 0.281 & 0.375 & 0.156 \\
 & Mean Estimate & 0.899 & 0.189 & 0.281 & 0.372 & 0.158 \\
 & s.d. & 0.011 & 0.015 & 0.010 & 0.019 & 0.004 \\
 &  &  &  &  &  &  \\
\multirow{3}{*}{$T_2$} & True & 0.900 & 0.188 & 0.281 & 0.375 & 0.156 \\
 & Mean Estimate & 0.900 & 0.180 & 0.286 & 0.378 & 0.156 \\
 & s.d. & 0.014 & 0.070 & 0.043 & 0.046 & 0.001 \\
 &  &  &  &  &  &  \\
\multirow{3}{*}{$T_3$} & True & 0.900 & 0.219 & 0.344 & 0.438 & - \\
 & Mean Estimate & 0.899 & 0.219 & 0.343 & 0.438 & - \\
 & s.d. & 0.011 & 0.000 & 0.010 & 0.010 & - \\
 &  &  &  &  &  &  \\ \bottomrule
\end{tabular}
\caption{The Table reports the results from the SFEM simulation for the low noise ($\beta=0.90$). $\pi^{k}$ is the mixing probability (proportion of subjects) of strategy $k$. \(gm\) stands for the G\&M type, \(alt\) for the altruist, \(coop\) for the conditional co-operator, and  \(free\) for the free rider. The frequency of types is set to 6 (0.188), 9 (0.281), 12 (0.375) and 5 (0.156) subjects for each type respectively, in $T_1$ and $T_2$, and 7 (0.219), 11 (0.344) and  14 (0.438) subjects in $T_3$ for the G\&M  the altruist, and the conditional co-operator.}
\label{tab:sim3}
\end{table}

From the tables, two notable points emerge. Firstly, the parameter $\beta$ is estimated with remarkable precision across all three levels of noise and treatments. Secondly, except for $T_1$ in the high noise specification, the frequencies of the types are recovered with exceptional precision, becoming nearly identical to the true values in the low noise specification.

Specifically, in $T_1$ under the high noise specification, although the ranking of the types is correctly recovered (conditional co-operators being the majority), the frequency of the G\&M type is overestimated (0.257 compared to the true 0.188), and that of the conditional co-operators is underestimated (0.312 instead of 0.375). However, for the other two types, the estimated frequencies are successfully recovered. Notably, this issue is absent in the other two treatments and completely disappears as the level of noise decreases.

The estimated $\beta$ across all treatments, using the actual experimental data, falls within the range of (0.82,0.89). This suggests that the results from the medium/low noise simulation are more relevant to our case. In summary, our simulations affirm the suitability of the SFEM methodology in identifying strategies and reliably recovering noise and frequency estimates, even when working with small sample sizes and situations where different models predict relatively similar decision patterns.

\newpage
\section{Regression Analysis} \label{sec:regression}
To obtain a better insight on the determinants of  contribution, in this appendix we report the results of regression analysis. Table \ref{tab:coefficients} reports the average marginal effects of Probit regressions. As dependent variable we use the binary decision to contribute or defect and we explore how the probability of contributing varies depending on whether the player is in position 2 (pos2) or in one of the uncertain positions 3 or 4 (pos34). We also examine the role of the information condition, namely whether the sample contained one contribution observation (cond1) or 2 (cond2). We included gender, degree and end-game effects (last), where we found no significance in any of the coefficients. Finally, to test the impact of learning in decisions, we include the round as a variable. 
\begin{table}[H]
\begin{center}
\begin{tabular}{l c c c}
\hline
 & $T_1$ & $T_2$ & $T_3$ \\
 \hline
& $Contribute$ & $Contribute$ & $Contribute$ \\
 \hline
round          & $-0.007$       & $-0.004$       & $-0.017^{*}$   \\
               & $(0.007)$      & $(0.007)$      & $(0.007)$      \\
pos2           & $-0.413^{***}$ & $-0.362^{***}$ & $-0.508^{***}$ \\
               & $(0.085)$      & $(0.084)$      & $(0.071)$      \\
pos34          & $-0.497^{***}$ & $-0.335^{***}$ & $-0.693^{***}$ \\
               & $(0.073)$      & $(0.070)$      & $(0.074)$      \\
cond1          & $0.387^{***}$  & $0.243^{***}$  & $0.442^{***}$  \\
               & $(0.063)$      & $(0.055)$      & $(0.062)$      \\
cond2          & $0.410^{***}$  &                & $0.551^{***}$  \\
               & $(0.067)$      &                & $(0.065)$      \\
\hline
Num. obs.      & $720$          & $560$          & $720$          \\
Log Likelihood & $-395.791$     & $-314.113$     & $-396.968$     \\
Deviance       & $791.582$      & $628.225$      & $793.936$      \\
AIC            & $803.582$      & $638.225$      & $805.936$      \\
BIC            & $831.058$      & $659.865$      & $833.411$      \\
\hline
\multicolumn{4}{l}{\scriptsize{$^{***}p<0.001$; $^{**}p<0.01$; $^{*}p<0.05$}}
\end{tabular}
\caption{Average marginal effects from Probit regressions with robust standard errors clustered on individuals with the decision to contribute being the dependent variable.}
\label{tab:coefficients}

\end{center}
\end{table}

The results are largely in line with the main analysis. The probability of contributing is higher when the sample is full co-operation (2,2), compared to observing only 1 co-operation (2,1) for both treatments 1 and 3. The position plays a role with the probability of defecting being higher when subjects are placed in positions 3 and 4 (face positional uncertainty) compared to 2. This is particularly the case in $T_3$ where subjects placed at the end of the sequence have an incentive to free-ride, where the probability of defection is almost 1.5 times as high, compared to $T_1$. There are no gender effects and there is no evidence of end-game effects which indicates that the subjects approached each round of the game as a one-shot game. Finally, the effect of round is not significant for Treatments 1 and 2, while subjects tend to decrease their contributions by approximately 0.02 during the session in Treatment 3. 
\end{appendices}
\end{document}